\documentclass[a4paper]{article}
\pdfoutput=1
\usepackage{jheppub}
\usepackage{comment}
\usepackage{url}
\usepackage{graphicx}
\usepackage{booktabs}
\usepackage{amsmath}
\usepackage{amssymb}
\usepackage{amsfonts}
\usepackage{hyperref}
\usepackage{appendix}
\usepackage{physics}
\usepackage[utf8]{inputenc}
\usepackage{bm}
\usepackage[normalem]{ulem}

\graphicspath{{figures/}}

\newcommand{\khat}{\hat{\bm{k}}_{\rm DM}}
\newcommand{\rhat}{\hat{\bm{r}}}
\newcommand{\Xhat}{\hat{\bm{X}}}
\newcommand{\thetahat}{\hat{\bm{\theta}}}
\newcommand{\phihat}{\hat{\bm{\varphi}}}
\newcommand{\uhat}{\hat{\bm{u}}}
\newcommand{\vhat}{\hat{\bm{v}}}
\newcommand{\nhat}{\hat{\bm{n}}}

\newcommand{\Rearth}{R_\oplus}

\newcommand{\NAB}{\mathcal{N}_{AB}}
\newcommand{\NbarAB}{\tilde{\mathcal{N}}_{AB}}
\newcommand{\xsat}{\hat X_{\rm sat}}

\newcommand\eea{\end{eqnarray}}
\newcommand\bea{\begin{eqnarray}}

\def\l{\left(}
\def\r{\right)}

\preprint{LITP-26-12}
\title{A directional force template for quadratically coupled ultralight dark matter} 

\author[a]{Dawid Brzeminski}
\author[a]{Aaron Pierce}

\affiliation[a]{Leinweber Institute for Theoretical Physics, Department of Physics,
University of Michigan, Ann Arbor, MI 48109, USA}

\emailAdd{dawid@umich.edu}
\emailAdd{atpierce@umich.edu}

\abstract{
Quadratic couplings between ultralight scalar dark matter and Standard Model fields can produce a distorted dark-matter field profile around the Earth. Gradients in the field induce a non-radial, composition-dependent force that can be suppressed at the Earth's surface while remaining accessible to space-based experiments. 
The MICROSCOPE satellite, which searched for violations of the equivalence principle, can constrain this force, but existing results assume a radial force, and they cannot be directly translated into an optimal bound in the anisotropic regime. 
We develop a signal template for this regime by 
organizing the force into radial and polar multipole coefficients and projecting the force onto the MICROSCOPE measurement axis. 
We use this template to recast the published MICROSCOPE constraint using the component of the signal that overlaps with the radial-force template. We estimate the sensitivity gain that would be provided by an analysis utilizing the additional non-overlapping signal.
Such an analysis  could improve  sensitivity to the couplings of quadratically coupled scalar dark matter by more than an order of magnitude relative 
to the radial-force recast 
for dark matter masses $\gtrsim 10^{-9}$ eV.}

\begin{document}
\emergencystretch=3em
\maketitle
\section{Introduction}
\label{sec:introduction}
 
Extensive laboratory, astrophysical, and cosmological searches for dark matter (DM) have yet to uncover a non-gravitational signal. In this work we focus on a class of ultralight scalar-DM models in which the absence of terrestrial signals can be a consequence of the environmental response of the DM field~\cite{Olive:2007aj,Arvanitaki:2014faa,Brzeminski:2020uhm,Banerjee:2022sqg,Delaunay:2025pho}. Crucially, these models can produce signals in space-based experiments.

If ultralight scalar dark matter possesses quadratic couplings to Standard Model (SM) fields, it will have an effective mass that depends on the ambient density and composition of ordinary matter~\cite{Olive:2007aj,Hees:2018fpg,Banerjee:2022sqg,Bouley:2022eer,Antypas:2022asj}. The propagation of a DM wave in the presence of a macroscopic body is effectively a scattering problem; a region where SM matter increases the effective DM mass acts as a potential barrier. This can attenuate the DM field inside a medium. For sufficiently large quadratic couplings, the Earth and/or its atmosphere can  screen terrestrial experiments from the scalar field.  Space-based experiments can retain sensitivity by probing the unscreened (exterior) field~\cite{Brzeminski:2026rox,Derevianko:2021kye,Schkolnik:2022utn,Budker:2024bzj}. The MICROSCOPE mission, which tested the weak equivalence principle with differential electrostatic accelerometers in low Earth orbit~\cite{Touboul:2017grn,MICROSCOPE:2022doy}, is a natural experiment with which to probe this DM field and the forces it can induce. 

Our focus is on the force induced by inhomogeneities in the background DM field profile generated by its interactions with  Earth.
The origin of this ``wake" or ``gradient" force was  described in Ref.~\cite{Hees:2018fpg} and further developed in Refs.~\cite{VanTilburg:2024xib, Barbosa:2024pkl,Gan:2025nlu}. The quadratic couplings of the DM to SM fields can induce a shift in the mass of a macroscopic body by changing  the mass parameters of its fundamental constituents and/or  their binding energies. In this way, the DM scalar profile acts as a composition-dependent potential, and the induced force is given by the gradient of the expectation value of this (spatially dependent) potential. MICROSCOPE can  constrain these forces through its differential acceleration measurements.

Previous works on quadratically coupled scalar dark matter used the spherically symmetric, low-momentum limit of the wake force~\cite{Hees:2018fpg,Banerjee:2022sqg} to recast the published MICROSCOPE bounds \cite{Touboul:2017grn,MICROSCOPE:2022doy}. This approximation is reliable when the de Broglie wavelength of the DM is longer than the size of the Earth, corresponding to $m_{\rm DM} \lesssim 10^{-11} \, \rm eV$ for virial DM velocities. At higher DM masses, finite-size effects become important. Combined with the preferred direction set by the DM wind, these effects generate an anisotropic profile, which impacts the nature of the force. Ref.~\cite{Gan:2025nlu} illustrated this anisotropy by comparing the force on opposite sides of the scattering body, while Ref.~\cite{Brzeminski:2026rox} developed a compact multipole description of the directional field profile.  
Another recent work~\cite{Fu:2026atp} discussed the ability to constrain the force from the dark matter wind at masses higher than those considered here. In that case, dark matter imparts its momentum directly to an object via scattering~\cite{Fukuda:2021drn, Day:2023mkb}. 
We briefly comment on this DM wind force in section~\ref{sec:WindComments}.  MICROSCOPE has also looked for the effects of Lorentz-violating backgrounds \cite{Pihan-LeBars:2019qsd}; these results are not immediately translatable to the current model.

The primary purpose of this work is to develop a  MICROSCOPE signal template for the wake force in the high $m_{\rm DM}$  (directional) regime.
Building on Ref.~\cite{Brzeminski:2026rox}, we derive a directional parametrization of the expectation value of the force described above  in terms of radial and polar force multipoles at MICROSCOPE altitudes.  The MICROSCOPE experiment has a preferred axis along which it performs a measurement.  As the satellite moves through space, the orientation of this axis changes.  The force multipole expansion provides an efficient formalism for projecting the DM-induced force onto the experiment's measurement axis. For a fixed $m_{\rm DM}$, the field-profile calculation determines the relative values of these force multipoles, so the resulting MICROSCOPE waveform is specified up to an overall normalization, apart from the usual nuisance parameters associated with the experimental analysis. 

MICROSCOPE searched for violations of the equivalence principle by looking for forces with a particular radial-force frequency
that is associated with a combination of the orbital and spin frequencies of the satellite \cite{MICROSCOPE:2022doy}. We demonstrate that the projected force does not reside exclusively in the radial-force frequency band that the standard MICROSCOPE EP analysis targets.  Rather, it populates additional ``sidebands" at different frequencies.  Neglecting these sidebands effectively discards some of the signal. In this work, we recast existing MICROSCOPE results using the part of the signal that overlaps with the usual radial-force frequency band, but also estimate the potential gain relative to this recast from a dedicated directional analysis. 

The paper is organized as follows. In section~\ref{sec:force_parametrization} we review quadratically coupled DM and set notation. We review the DM field profile in the strong-coupling regime and derive the resulting acceleration from the wake force. In section~\ref{sec:microscope_template} we translate this acceleration into a MICROSCOPE signal template. We derive the frequency structure of the signal, recast the existing MICROSCOPE constraint, and estimate the sensitivity improvement possible in a dedicated multi-frequency analysis. We conclude in section~\ref{sec:outlook}.  Our baseline analysis neglects the effects of the atmosphere and simply treats the Earth as a hard sphere.  We assess the robustness of this ``hard-surface baseline" approximation  in an appendix, where we solve the scalar propagation problem in the presence of a spherically averaged atmospheric potential. 

\section{Force parametrization}
\label{sec:force_parametrization}

The goal of this section is to provide a parametrization of the DM-induced wake force relevant for MICROSCOPE.  To that end, we first provide an overview of the dark matter model and its couplings to matter.  We  discuss how interactions with Earth set up a nontrivial profile for the DM field.  We then review how to compute this profile, and from that profile we calculate the induced force.

\subsection{Quadratic couplings and macroscopic bodies}
\label{subsec:lagrangian}

We follow the low-energy dilaton-like convention used in Refs.~\cite{Damour:2010rp,Damour:2010rm, Arvanitaki:2014faa,Banerjee:2022sqg}. The quadratic interactions of a scalar $\phi$ with Standard Model fields may be written as
\begin{equation}
\label{eq:dilaton_like}
\begin{aligned}
\mathcal{L} \supset \frac{\kappa^2 \phi^2}{2}\Big[ \frac{d_e^{(2)} }{4 e^2} F_{\mu\nu}F^{\mu\nu}
-\frac{d_g^{(2)}\beta_3}{2g_3}G^A_{\mu\nu}G^{A \mu\nu}
-d_{m_e}^{(2)}m_e \bar{\psi}_e\psi_e \\
-\sum_{i=u,d}{(d_{m_i}^{(2)}+\gamma_{m_i} d_g^{(2)})m_i\bar{\psi}_i \psi_i} \Big]\; .
\end{aligned}
\end{equation}
Here $\kappa=\sqrt{4\pi G_N}$, with $G_N$ Newton's gravitational constant, is used as an overall normalization. The quantities $e$ and $g_3$ are the electromagnetic and QCD gauge couplings, $\beta_3 \equiv \partial g_3/\partial\log\mu$ is the QCD beta function, $F_{\mu\nu}$ and $G^A_{\mu\nu}$ are the photon and gluon field strengths, $m_e$ and $m_i$ are the electron and light-quark masses, and $\gamma_{m_i}=-\partial\log m_i/\partial\log\mu$ is the anomalous dimension of the mass. The coefficients $d_X^{(2)}$ with $X\in\{e,g,m_e,m_u,m_d\}$ parametrize quadratic couplings to the corresponding low-energy parameter.  The dimensionless charge $Q_X^{(A)}$ is the logarithmic response of the mass of body $A$ to $X$, $Q_X^{(A)}=\partial\log m_A/\partial\log X$, evaluated for the body's composition; see Refs.~\cite{Damour:2010rp,Damour:2010rm}.  Unless otherwise stated, the force expressions below are written for one nonzero $d_X^{(2)}$ at a time; for simultaneous couplings, the combination $d_X^{(2)}Q_X^{(A)}$ should be replaced by the corresponding sum over $X$. Instead of quoting separate up- and down-quark mass couplings, it is conventional to use the average quark mass $\hat{m}\equiv(m_u+m_d)/2$ and the quark-mass difference $\delta m\equiv m_d-m_u$. The corresponding quadratic couplings are
\begin{equation}
d_{\hat{m}}^{(2)}
\equiv
\frac{m_u d_{m_u}^{(2)}+m_d d_{m_d}^{(2)}}{m_u+m_d},
\qquad
d_{\delta m}^{(2)}
\equiv
\frac{m_d d_{m_d}^{(2)}-m_u d_{m_u}^{(2)}}{m_d-m_u}.
\label{eq:mhat_deltam_defs}
\end{equation}
When this quark-mass basis is used, the coupling label $X$ may equivalently run over $X\in\{g,e,m_e,\hat{m},\delta m\}$.

The quadratic nature of the couplings in eq.~\eqref{eq:dilaton_like} has consequences for both ordinary matter and the dark matter. For Standard Model parameters $X$,
\begin{equation}
\frac{\Delta X(\bm{r},t)}{X}
=
\frac{d_X^{(2)}\kappa^2}{2}
\phi^2(\bm{r},t).
\label{eq:dxoverx}
\end{equation}
Changes in fundamental parameters affect constituent masses and binding energies, so macroscopic masses inherit the field dependence. The response of an object to changes in fundamental parameters depends on the composition of the body. For a body $A$,
\begin{equation}
m_A(\bm{r},t)
=
m_{A,0}
\left[
1+\frac{d_X^{(2)}\kappa^2}{2}
Q_X^{(A)}
\phi^2(\bm{r},t)
\right].
\label{eq:mass_shift}
\end{equation}
Because this field-dependent mass acts as a potential energy, a spatial variation in the field $\phi$ results in a composition-dependent acceleration of the object. 

The same interaction also changes the scalar dark-matter mass in the presence of matter. In a medium,
\begin{equation}
m_{\rm DM}^2(\bm{r})
=
m_{\rm DM}^2
+
\sum_X d_X^{(2)}Q_X(\bm{r})\kappa^2\rho_{\rm SM}(\bm{r})
\equiv
m_{\rm DM}^2+\Delta m_{\rm DM}^2(\bm{r}),
\label{eq:DM_mass}
\end{equation}
where $Q_X(\bm{r})$ is the local composition-weighted dilatonic charge of the medium and $\rho_{\rm SM}$ is its Standard Model energy density.  In this work we focus on $\Delta m_{\rm DM}^2>0$, for which matter acts as a potential barrier and can attenuate the field. For $\Delta m_{\rm DM}^2<0$, matter instead lowers the effective mass squared, which can enhance the local field and its gradients and, when sufficiently strong, lead to a tachyonic instability and scalarization.\footnote{Negative matter-induced mass shifts have been studied in several related contexts. White-dwarf structure can constrain scalarization driven by quadratic matter couplings, including exceptionally light QCD-axion scenarios~\cite{Bartnick:2025lbg,Bartnick:2025lbv}. For axion dark matter, Earth-induced enhancement of the field and its gradients can modify standard axion observables~\cite{Banerjee:2025dlo,Huang:2026tgv}, while Ref.~\cite{Gue:2025nxq} performed a low-momentum MICROSCOPE analysis of the associated composition-dependent force.}  When $\Delta m_{\rm DM}^2\gg k_{\rm DM}^2$, with $k_{\rm DM}\sim 10^{-3}m_{\rm DM}$ the characteristic incoming momentum, the Earth or atmosphere strongly attenuates the field. In the low-momentum limit $k_{\rm DM}\Rearth\ll 1$, corresponding to $m_{\rm DM} \ll 3 \times 10^{-11}$ eV, the Earth is approximately point-like relative to the de Broglie wavelength, and the resulting force is radial~\cite{Hees:2018fpg}. When $k_{\rm DM}\Rearth\gtrsim 1$, this approximation fails. Characterizing the non-radial force present in this regime and translating it into MICROSCOPE observables is the main focus of this work. For this purpose, in section~\ref{subsec:field_profile} we review the field-profile description introduced in Ref.~\cite{Brzeminski:2026rox}, which serves as the starting point for the wake force description.

Before turning to the field description, we briefly comment on the observable. In the regime of interest, roughly $m_{\rm DM}\gtrsim 10^{-12}\,{\rm eV}$, the  oscillation frequency set by the mass is much faster than the MICROSCOPE sampling rate $\Omega_{\rm samp}/(2\pi)=4\,{\rm Hz}$. The relevant signal is therefore not the phase-resolved force, but its average over the rapid oscillations and over the stochastic phases of the DM wave.
For a homogeneous field in vacuum, this averaged gradient force would vanish: the time-averaged intensity $\langle \phi^2\rangle$ is spatially uniform, so $\langle \phi\nabla\phi\rangle=\nabla\langle\phi^2\rangle/2=0$.
The situation is different near the Earth. The Earth distorts the incoming DM wave and produces a spatially dependent, anisotropic intensity profile, so
$\langle \phi\nabla\phi\rangle \neq 0.$
When we refer to the  
force below, we mean the expectation value of the force sourced by the Earth-distorted DM profile.

\subsection{Review of the field profile induced by the Earth in the large-coupling regime}
\label{subsec:field_profile}
Following Ref.~\cite{Brzeminski:2026rox}, we discuss the spatial dependence of the dark matter profile.  This provides the input needed to compute the force probed by MICROSCOPE.
The behavior of the scalar is described by the Klein-Gordon equation
\begin{equation}
\left(\Box+m_{\rm DM}^2(\bm{r})\right)\phi(\bm{r},t)=0.
\label{eq:KG}
\end{equation}
In the large-coupling hierarchy
\begin{equation}
k_{\rm DM}^2 \ll \Delta m_{{\rm DM},{\rm atm}}^2
\ll
\Delta m_{{\rm DM},\oplus}^2 ,
\label{eq:dirichlet_regime}
\end{equation}
the field is strongly screened by the atmosphere and Earth, and the exterior problem is well-approximated by a Dirichlet boundary condition at the scattering surface. In the formulae below, we adopt a hard-sphere approximation in which this surface is placed at $\Rearth$. Thus, $\xi \equiv \vert \bm{r} \vert/\Rearth$ and $k\Rearth$ are the relevant dimensionless variables. At MICROSCOPE altitudes the Earth can be treated as spherical for this purpose.

Assuming that the dark matter is described by an isothermal halo, the exterior solution for $\phi$ can be constructed by forming a superposition of scattering solutions with incoming momenta drawn from the boosted Maxwell-Boltzmann (BMB) distribution,
\begin{equation}
\phi(\bm{r},t)
=
\int d^3 k\,\sqrt{f_{\rm BMB}(\bm{k})}\,
\tilde{\phi}(\bm{r},t;\bm{k}),
\qquad
f_{\rm BMB}(\bm{k})
=
\left(\frac{1}{2\pi\sigma_k^2}\right)^{3/2}
\exp\!\left[-\frac{(\bm{k}-\bm{k}_{\rm DM})^2}{2\sigma_k^2}\right],
\label{eq:BoostedMB}
\end{equation}
where $\bm{k}_{\rm DM}$ is the mean dark-matter wind momentum and $\sigma_k\simeq k_{\rm DM}/\sqrt{2}$. We write
\begin{equation}
\label{eq:psidef}
\tilde{\phi}(\bm{r},t;\bm{k})
=
{\rm Re}\!\left[
e^{i\omega t+i\chi_{\bm{k}}}\psi(\bm{r};\bm{k})
\right],
\qquad
\omega^2=k^2+m_{\rm DM}^2,
\end{equation}
where $\chi_{\bm{k}}$ is an independently drawn random phase for each momentum $\bm{k}$.
Imposing the Dirichlet boundary condition, the exterior solution for a single incoming momentum may be written as
\begin{equation}
\psi
=
\abs{\psi_0}
\sum_{\ell}(2\ell+1)i^\ell
P_\ell(\cos\theta)
\left[
j_\ell(kr)
-
\frac{j_\ell(k\Rearth)}{h_\ell(k\Rearth)}
h_\ell(kr)
\right],
\qquad r\geq \Rearth .
\label{eq:multipole}
\end{equation}
Here $j_\ell$ and $h_\ell$ are spherical Bessel and Hankel functions, with 
$h_{\ell} \equiv j_{\ell} + i n_{\ell}$ and $n_{\ell}$ the spherical Bessel function of the second kind.
We define $\theta$ as the angle between $\bm{r}$ and $\bm{k}$, and $\psi_0$ is the amplitude of the incoming plane-wave. Choosing the polar axis to align with the mean wind direction $\bm{k}_{\rm DM}$, the cosine of the angle between a momentum direction $\hat{\bm{k}}$ and the position direction  $\rhat$ is
\begin{equation}
\cos\theta
=
\hat{\bm{k}}\cdot\rhat
=
\sin\theta_k\cos\varphi_k\sin\theta_r
+
\cos\theta_k\cos\theta_r ,
\label{eq:dotproduct}
\end{equation}
where $\theta_k$ and $\theta_r$ are measured with respect to $\bm{k}_{\rm DM}$ (i.e. the polar axis/mean DM wind direction).  Azimuthal symmetry allows the choice $\varphi_r=0$.

\begin{figure}[t]
\centering
\includegraphics[width=0.70\linewidth]{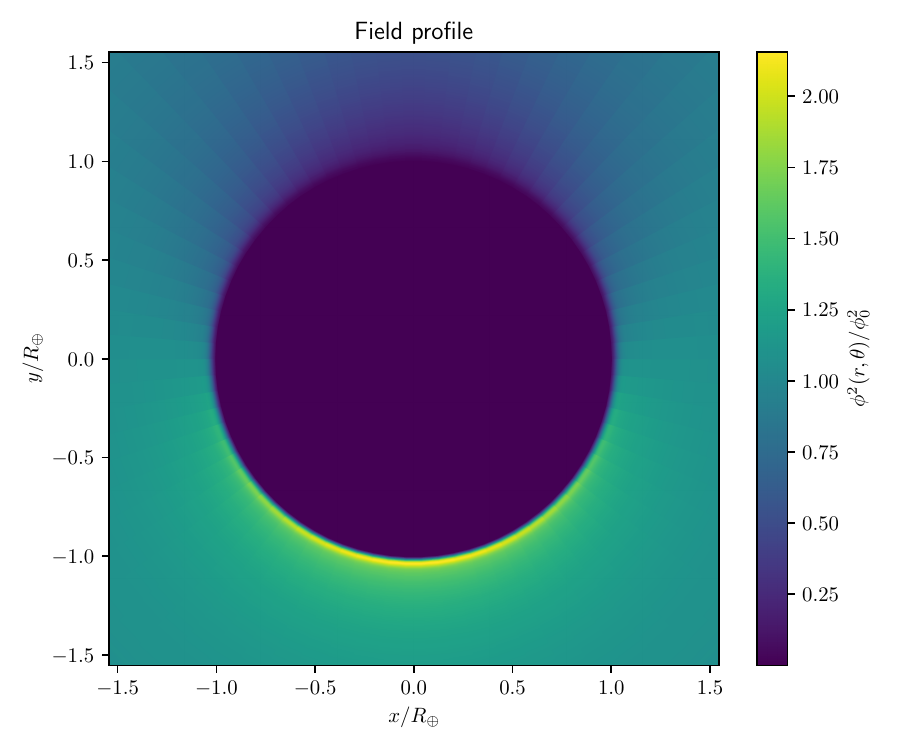}
\caption{Example scalar-density profile in the large-coupling regime for $m_{\rm DM} \approx 10^{-9}$ eV,  following Ref.~\cite{Brzeminski:2026rox}. The incoming dark-matter wind selects a preferred direction, so for $k_{\rm DM}\Rearth\gtrsim 1$ the field profile is not spherically symmetric. The DM wind is incident from the bottom of the figure. The color bar indicates the local value of $\phi^2$ relative to the asymptotic reference value 
$\phi^2_0$.}
\label{fig:field_profile}
\end{figure}

Averaging over time and over the random phases $\chi_{\bm{k}}$ leaves
\begin{equation}
\left\langle\phi^2(\bm{r},t)\right\rangle
=
\int d^3k\, f_{\rm BMB}(\bm{k})\,\abs{\psi(\bm{r};\bm{k})}^2 .
\label{eq:averagedphi}
\end{equation}
The resulting profile has the familiar wake/shadow structure of a wave scattered from an opaque body, as illustrated in figure~\ref{fig:field_profile}.

The profile can 
be organized into Legendre multipoles \cite{Brzeminski:2026rox},
\begin{equation}
\left\langle \phi^2(\bm{r}) \right\rangle
=
\frac{\rho_{\rm DM}}{m_{\rm DM}^2}
\sum_{\ell=0}^{\infty}
a_\ell(\xi,m_{\rm DM})P_\ell(\cos\theta_r).
\label{eq:phi2_profile}
\end{equation}
When the mass dependence is clear, we suppress it and write $a_\ell(\xi)$.  The monopole coefficient $a_0$ describes the angular average of the profile, while $a_1$ is the contribution that gives the leading nontrivial spatial dependence; it becomes important once the de Broglie wavelength is shorter than the Earth's radius. In practice, keeping multipoles up to $\ell=4$ is sufficient to characterize the field profile at the $1\%$ level over the parameter range considered in this work~\cite{Brzeminski:2026rox}.

\subsection{Expectation value of the wake force}
\label{subsec:force_expectation}
We now discuss how this profile leads to a ``wake" or ``gradient" force.
The DM field profile induces an effective composition-dependent potential, whose gradient produces an acceleration that can be probed by MICROSCOPE. Starting with the expression for the field profile in eq.~\eqref{eq:phi2_profile}, our task is to derive the induced acceleration in terms of the $a_{\ell}$ coefficients. 

Dividing by the body mass, we can write the composition-dependent fractional mass shift of a body $A$ as
\begin{equation}
\frac{\left\langle \delta m_A(\bm{r})\right\rangle}{m_{A,0}}
=
\mathcal{N}_A
\sum_{\ell=0}^{\infty}
a_\ell(\xi,m_{\rm DM})P_\ell(\cos\theta_r),
\label{eq:potential_expansion}
\end{equation}
where
\begin{equation}
\mathcal{N}_A
=
\frac{1}{2}d_X^{(2)}\kappa^2 Q_X^{(A)}
\frac{\rho_{\rm DM}}{m_{\rm DM}^2}.
\label{eq:NA_def}
\end{equation}
Taking the gradient of the potential, we  arrive at an expression for the acceleration:
\begin{equation}
\bm{a}_A(\xi,\theta_r)
=
-\frac{\mathcal{N}_A}{\Rearth}
\left[
\rhat \sum_{\ell=0}^{\infty}a_\ell'(\xi)P_\ell(\cos\theta_r)
+
\thetahat\frac{1}{\xi}
\sum_{\ell=1}^{\infty}a_\ell(\xi)P_\ell^1(\cos\theta_r)
\right],
\label{eq:force_master}
\end{equation}
where the prime 
denotes $d/d\xi$, and
\begin{equation}
P_\ell^1(\cos\theta_r)
\equiv
\frac{\partial}{\partial\theta_r}P_\ell(\cos\theta_r)
\label{eq:associated_legendre_convention_force}
\end{equation}
defines the first-order associated Legendre polynomial of degree $\ell$ used throughout. Note that ${\bf \hat{\phi}}$ does not appear in eq.~(\ref{eq:force_master}) owing to the azimuthal symmetry.  It is useful to define radial ($b_{\ell}$) and polar ($t_{\ell}$) force coefficients in terms of the multipoles that describe the field profile as
\begin{equation}
b_\ell(\xi)=-a_\ell'(\xi),
\qquad
t_\ell(\xi)=-\frac{a_\ell(\xi)}{\xi}.
\label{eq:bt_raw_defs}
\end{equation}
Then the acceleration can be written as
\begin{equation}
\bm{a}_A(\xi,\theta_r)
=
\frac{\mathcal{N}_A}{\Rearth}
\left[
\rhat\sum_{\ell=0}^{\infty}b_\ell(\xi)P_\ell(\cos\theta_r)
+
\thetahat\sum_{\ell=1}^{\infty}t_\ell(\xi)P_\ell^1(\cos\theta_r)
\right].
\label{eq:force_bt}
\end{equation}
The above parametrization allows us to assess the relative contribution of each multipole coefficient to the force. The $b_0$ coefficient represents a purely radial contribution from the isotropic (monopole) part of the field profile ($t_0$ does not contribute since $P_0^1 = 0$). In contrast, the $b_1$ and $t_1$ coefficients are the leading contributions from an anisotropic part of the potential. 
\begin{figure}[t]
\centering
\includegraphics[width=0.49\linewidth]{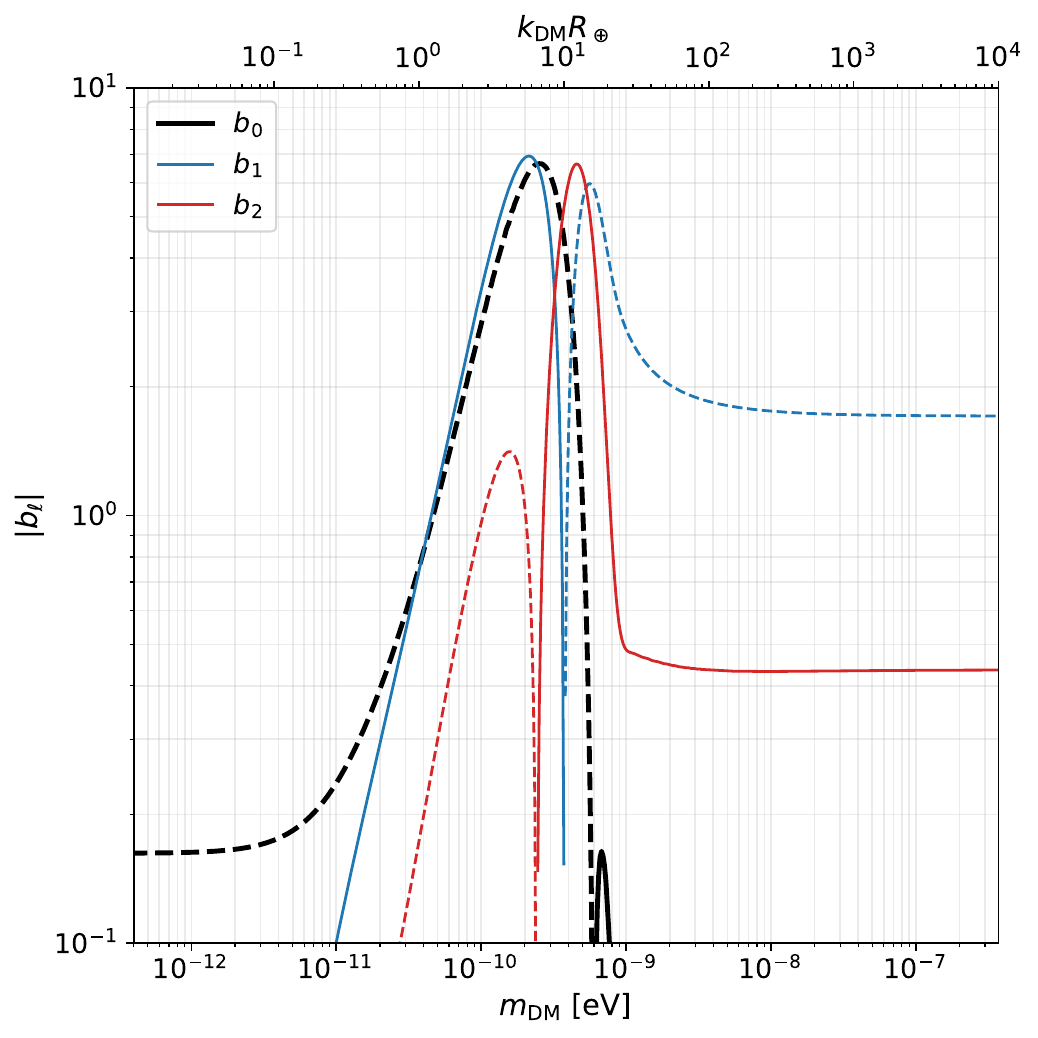}
\includegraphics[width=0.49\linewidth]{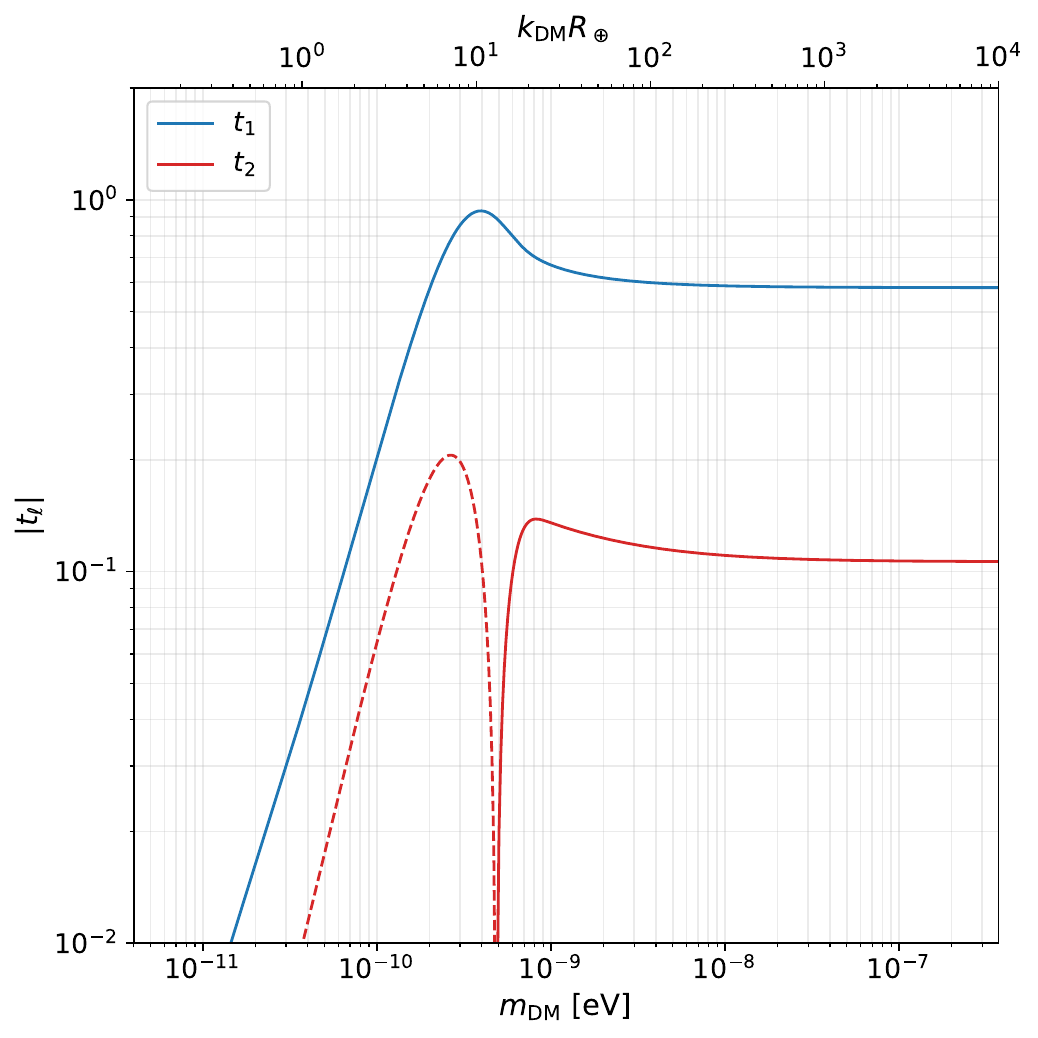}
\caption{Radial $b_\ell$ (left panel) and polar $t_\ell$ (right panel) force coefficients evaluated at $\xi = 1.11$, a $\xi$ value corresponding to the altitude of MICROSCOPE. The lower and upper horizontal axes show $m_{\rm DM}$ and $k_{\rm DM}\Rearth$, respectively (assuming $k_{\rm DM}\approx 10^{-3} m_{\rm DM}$), while the vertical axis shows the absolute value of the corresponding coefficient. Solid (dashed) segments denote positive (negative) coefficient values, with the multipole order indicated by the legend. }
\label{fig:b_coefficients}
\end{figure}

In figure~\ref{fig:b_coefficients} we demonstrate how these coefficients depend on dark matter mass $m_{\rm DM}$ in the large-coupling limit, evaluated at $\xi = 1.11$ (corresponding to the MICROSCOPE altitude).  The lowest values of $\ell$ typically dominate the signals of interest.  Details on how to calculate the $b_\ell$ and $t_\ell$ coefficients are given in appendix~\ref{app:master_integrals}.  In this figure we represent positive values of coefficients with solid lines and negative values with dashed lines of the same color. As expected, in the low-mass limit $(m_{\rm DM} < 10^{-11}\, \rm eV)$, where we expect the isotropic approximation to be valid, the force is dominated by $b_0$, which asymptotes to 
\begin{equation}
b_{0,{\rm LM}}(\xi)
=
-2\xi^{-2}\left(1-\frac{1}{\xi}\right).
\label{eq:b0_lm_raw}
\end{equation}
The limit is consistent with $a_0$ reaching its asymptotic value, $a_{0, {\rm LM}} = (1-1/\xi)^2$, see Ref.~\cite{Brzeminski:2026rox}.
In the high-mass limit, the contribution from $b_0$ is negligible as $a_0 \approx 1$, approximately independent of  altitude. The leading contribution to the gradient comes from $b_1$ and $t_1$, which is consistent with the anisotropy of the potential depending on the distance to the surface as shown in Ref.~\cite{Brzeminski:2026rox}. 
  As we will discuss, $b_2$ can provide an important correction to the radial force.  From the figure, we can anticipate that this correction will be particularly important for DM masses $m_{\rm DM} \gtrsim 10^{-9}$ eV. 

The main lesson from figure~\ref{fig:b_coefficients} is that the force profile becomes anisotropic once the low-momentum, isotropic approximation for the profile breaks down. This anisotropy is not simply equivalent to an acceleration pointing along the DM wind. Since $\khat = \cos \theta_r \, \rhat - \sin \theta_r \, \thetahat$, and since $b_1 \neq t_1$, as seen in figure~\ref{fig:b_coefficients}, the non-radial correction to the acceleration has a direction that depends on the relative size of the radial and polar components. 

\subsection{Comments on a dark matter wind force}
\label{sec:WindComments}
As mentioned in the introduction, dark matter can directly impart momentum to a target. This is sometimes called  the wind force \cite{Day:2023mkb,Fukuda:2018omk}. Coherent scattering of dark matter from macroscopic targets has also been studied with torsion-balance experiments~\cite{Luo:2024ocg, Matsumoto:2025rcz}. Recently, this effect was discussed in the context of MICROSCOPE~\cite{Fu:2026atp,Liu:2025jlx}. If DM couples universally to nucleons, for masses $\lesssim 10^{-2}$ eV the long de Broglie wavelength $\lambda_{\rm dB}$ of the dark matter results in important interference effects between scattering from the two MICROSCOPE test masses, causing a substantial degradation in sensitivity~\cite{Fu:2026atp}. The wind force in this regime is suppressed by $(R/\lambda_{\rm dB})^2$, where $R$ is the separation of the test masses. For the dark matter considered here, however, the differing dilaton charges of the test masses cause an incomplete cancellation. In principle, a force proportional to $Q_A - Q_B$ but unsuppressed by the de Broglie wavelength remains.   
Nevertheless, once we account for interference between the test masses as well as screening from the MICROSCOPE satellite itself, we find that the wind-induced force is irrelevant for MICROSCOPE in the mass regime we consider.  
We do not consider it further in this work. A corresponding template could also be developed along the lines of the wake-force template derived here.

\section{MICROSCOPE signal template}
\label{sec:microscope_template}

We now translate the force parametrization into a MICROSCOPE signal template. MICROSCOPE searched for violations of the weak equivalence principle by measuring the differential acceleration of two concentric test masses aboard a low-Earth orbit (LEO) satellite~\cite{Touboul:2017grn,MICROSCOPE:2022doy}. In the standard equivalence principle (EP) interpretation of their data, the anomalous force is assumed to be radial, because it is sourced by the Earth. 

The experiment is designed so that  the satellite's spin, together with the orbital motion, modulates this radial signal into a particular location in frequency space. 
The DM-induced force considered here has a richer  structure. The Earth distorts the incident DM field, and the resulting force depends not only on the radial direction $\hat{\bm r}$, but also on the mean DM-wind direction $\khat$. Consequently, the projected acceleration along the MICROSCOPE sensitive axis contains power at frequencies 
beyond the usual radial-force band. We first derive the projected template in terms of the satellite attitude and the DM-wind geometry, then use a circular-orbit model to exhibit the frequency structure explicitly. We distinguish the signal component constrained by the published MICROSCOPE result from the additional components that could be used in a dedicated multi-harmonic analysis.

\subsection{Geometry and projected template}
\label{subsec:microscope_geometry}
MICROSCOPE measures the differential acceleration  projected along a sensitive instrument axis. The satellite spins at an angular frequency $\Omega_{\rm spin}$, which modulates an equivalence-principle violating signal away from the lowest-frequency noise and helps separate it from several systematic effects. In the simplified geometry below, the radial-force signal appears as a line in frequency space at 
\begin{equation}
    \Omega_{\rm EP} \equiv \Omega_{\rm spin}+\Omega_{\rm orb}, 
    \label{eq:OmegaEPDef}
\end{equation}
    where $\Omega_{\rm orb}$ is the angular frequency of the orbit, and $\Omega_{\rm spin} \gg \Omega_{\rm orb}$. 

For the dark matter-induced force, the differential acceleration for two materials $A$ and $B$ is
\begin{equation}
\Delta\bm{a}_{AB}
=
\frac{\mathcal{N}_{AB}}{R_\oplus}
\left[
\rhat\sum_{\ell=0}^{\infty}b_\ell(\xi)P_\ell(\mu)
+
\thetahat\sum_{\ell=1}^{\infty}t_\ell(\xi)P_\ell^1(\mu)
\right],
\label{eq:delta_a}
\end{equation}
where
\begin{equation}
\mathcal{N}_{AB}
=
\frac{\rho_{\rm DM}}{2m_{\rm DM}^2}
d\,\kappa^2\Delta Q^{AB}, \quad \mu \equiv \cos \theta_r.
\label{eq:delta_a_prefactor}
\end{equation}
We have simplified the notation for the coupling $d$, omitting sub- and super-scripts for brevity. For the remainder of the text we refer to the coupling as $d$ whenever the discussion applies to all coupling types, only restoring the full notation when we specify the interaction.
Here, $\Delta Q^{AB} \equiv Q^{(A)}-Q^{(B)}$ is
the difference in the dilatonic charges between the test masses. 
One is composed of a platinum/rhodium alloy and the other a titanium/aluminum/vanadium alloy.  The corresponding differences in the relevant dilatonic charges for $(e,m_{e}, g, \hat{m}, \delta m)$, are given by \cite{Hees:2018fpg} 
\begin{equation}
\label{eq:deltaQ}
\Delta Q^{AB}\simeq 10^{-3} (-1.9, 0.031, 4.6, -2.7, -0.19). 
\end{equation}
See appendix \ref{app:dilaton_charges} for general formulas for the dilatonic charges for elements with atomic number $Z$ and mass number $A$.

The experiment is sensitive along a single measurement axis $\bm{\hat{X}}(t)$. The experiment effectively records the projection of an anomalous acceleration along this sensitive direction:
\begin{equation}
\Gamma_X^{\rm DM}(t)
=
\Xhat(t)\cdot\Delta\bm{a}_{AB}(t)
=
\frac{\NAB}{\Rearth}
\left[
\left(\bm{\hat{X}}(t) \cdot\rhat(t)\right)\sum_{\ell=0}^{\infty}b_\ell(\xi)P_\ell(\mu(t))
+
\left(\Xhat(t)\cdot\thetahat(t)\right)
\sum_{\ell=1}^{\infty}t_\ell(\xi)P_\ell^1(\mu(t))
\right].
\label{eq:Gamma_X}
\end{equation}
Since the unit vector $\thetahat$ lies in the plane spanned by $\rhat$ and $\khat$, the above expression can be written in terms of these vectors. With the associated-Legendre definition of eq.~\eqref{eq:associated_legendre_convention_force},
\begin{equation}
\thetahat=\frac{\mu\rhat-\khat}{\sqrt{1-\mu^2}},
\qquad
P_\ell^1(\mu)\thetahat
=
\frac{\partial P_\ell(\mu)}{\partial \mu} \left(\khat-\mu\rhat\right),
\label{eq:theta_identity}
\end{equation}
we can write 
\begin{equation}
\Gamma_X^{\rm DM}(t) = 
\frac{\mathcal{N}_{AB}}{R_\oplus}
\left\lbrace b_0 X_r+\sum_{\ell=1}^{\infty}\left[
\left(
b_\ell P_\ell(\mu)
-
t_\ell\mu \frac{\partial P_\ell(\mu)}{\partial \mu}
\right)X_r
+
t_\ell \frac{\partial P_\ell(\mu)}{\partial \mu} X_k
\right]\right\rbrace,
\label{eq:Gamma_X_template}
\end{equation}
where we defined
\begin{equation}
X_r=\Xhat\cdot\rhat,
\qquad
X_k=\Xhat\cdot\khat .
\label{eq:XrXk}
\end{equation}
$X_{r}$ indicates the orientation of the sensitive axis with respect to the radial direction, and $X_{k}$ gives information about the orientation of the sensitive axis with respect to the mean DM wind direction. 
This form is useful for data analysis: the modulation of the signal is controlled by three time-dependent quantities,  $X_r(t)$, $X_k(t)$, and  $\mu(t)$. The first can be constructed purely from satellite attitude and gravitational-acceleration data. The remaining two  require the direction of  $\khat$ relative to the satellite. This direction can be inferred by combining satellite timing data with the Solar System velocity in the Galactic frame and the Earth’s orbital ephemeris.

Although the above signal model contains many $b_\ell$ and $t_\ell$ parameters,  
for a fixed DM mass, the force model presented in the previous section predicts the relative values of the $b_\ell$ and $t_\ell$ coefficients. Therefore, the template is specified up to an overall normalization, apart from the usual nuisance parameters associated with the experimental analysis. An efficient search strategy is then to divide the DM parameter space into mass bins and, within each bin, fix the relative values of the force multipoles. Depending on the DM mass being probed, different features of the signal dominate. 
In the remainder of this section, we explore these features using a semi-analytic approach and discuss how they can be exploited to improve the sensitivity of a MICROSCOPE analysis.

\subsection{Circular-orbit sideband structure}
\label{subsec:sidebands}

The purpose of this subsection is to show how the non-radial force enters the data. Throughout the mission, the satellite spin axis was approximately parallel to the orbital axis, and the orbit was Sun-synchronous, meaning that the orbital precession period was equal to one year. Since the orbit's eccentricity was reported to be less than $e<5 \times 10^{-3}$~\cite{MICROSCOPEDataPackageDescription}, the orbit was circular to a good approximation. We introduce an orbital basis $(\uhat,\vhat,\nhat)$, where $\nhat$ is the orbital angular-momentum direction and $\uhat,\vhat$ span the orbital plane. Let $\phi(t)$ be the orbital phase measured from $\uhat$. In the circular-orbit approximation, the satellite position $\rhat(t)$ is given by
\begin{equation}
\rhat(t)=\cos\phi(t)\uhat+\sin\phi(t)\vhat .
\label{eq:rhat_orbit}
\end{equation}
In this basis the mean dark-matter wind direction is decomposed as
\begin{equation}
\khat=w_u\uhat+w_v\vhat+w_n\nhat.
\label{eq:khat_orbital_components}
\end{equation}
We introduce an angle $\chi_{w}$ via 
\begin{equation}
W=\sqrt{w_u^2+w_v^2},
\qquad
w_u=W\cos\chi_w,\qquad
w_v=W\sin\chi_w .
\label{eq:chiangle}
\end{equation}
Here, $W$ is the magnitude of the dark-matter-wind projected onto the orbital plane, and $\chi_w$ is its azimuth in that plane.
Then
\begin{equation}
\mu(t)=W\cos\!\left[\phi(t)-\chi_w\right].
\label{eq:mu_orbital}
\end{equation}
Orbital precession and seasonal modulation make $W$ and $\chi_w$ slowly time-dependent in a full analysis. For now, we treat them as fixed.  However, as we will see in detail below, the value of $W$ is relevant  for the sensitivity of the experiment; it controls the projection of the non-radial part of the force onto the measurement/orbital plane.

To model the satellite spin, we let $\psi(t)$ be the spin phase of the sensitive axis and take
\begin{equation}
\Xhat(t)=\cos\psi(t)\uhat-\sin\psi(t)\vhat,
\label{eq:Xhat_spin}
\end{equation}
where the relative minus sign between $\vhat$ in eq. \eqref{eq:rhat_orbit} and eq. \eqref{eq:Xhat_spin}  indicates that the orbital and spin axes are anti-aligned, i.e. the spin axis is $-\hat{\bf n}$. In that convention $\phi(t) = \Omega_{\rm orb} t + \phi_0$ and $\psi(t) = \Omega_{\rm spin} t + \psi_0$. Note that the sensitive axis is orthogonal to ${\bf \hat{n}}$. We can write
\begin{equation}
X_r=\cos(\phi(t)+\psi(t)),
\qquad
X_k=W\cos(\psi(t)+\chi_w).
\label{eq:xr_xk_harmonic}
\end{equation}
Defining $a\equiv\phi-\chi_w$ and $s\equiv\psi+\chi_w$ and combining  Eqs.~(\ref{eq:mu_orbital}), (\ref{eq:xr_xk_harmonic})  with the signal template from eq. \eqref{eq:Gamma_X_template}, we can see the emerging harmonic structure in the first few multipoles:
\begin{align}
\ell=0:\quad
& b_0\cos(s+a),
\label{eq:harmonic_l0}\\
\ell=1:\quad
& \frac{W}{2}(b_1+t_1)\cos s
+
\frac{W}{2}(b_1-t_1)\cos(s+2a),
\label{eq:harmonic_l1}\\
\ell=2:\quad
&
\frac{3W^2}{8}(b_2+2t_2)\cos(s-a)
+
\frac{b_2}{4}(3W^2-2)\cos(s+a)
\nonumber\\
&\quad
+
\frac{3W^2}{8}(b_2-2t_2)\cos(s+3a).
\label{eq:harmonic_l2}
\end{align}
Since $d(s+q a)/dt = \Omega_{\rm spin} + q \, \Omega_{\rm orb}$, the monopole contributes to the usual radial-force band $\Omega_{\rm EP}$, see eq.~(\ref{eq:OmegaEPDef}). However, it is not the only contribution to that band. Even multipoles also contribute.  For example, the second term in the $\ell = 2$ expression provides a correction. In the high-mass (directional) regime, the dipole ($\ell=1$) term populates frequencies that differ from the radial band by $\pm \Omega_{\rm orb}$, i.e. $\Omega_{\rm spin}$ and $\Omega_{\rm spin}+ 2\,\Omega_{\rm orb}$, with amplitudes controlled by $b_1+t_1$ and $b_1-t_1$. Higher multipoles populate additional frequencies, again differing from the radial frequency by multiples of $\Omega_{\rm orb}$. We refer to these frequencies where some of the dark matter-induced force resides as \emph{sidebands}.
An analysis restricted to the radial-force band alone does not use all of the modeled directional signal.

The signal  structure can be organized compactly as
\begin{equation}
\Gamma_X^{\rm DM}(t)
=
\frac{\NAB}{\Rearth}
\sum_q A_q\cos(s+qa),
\label{eq:aq_expansion}
\end{equation}
where the amplitudes in the radial-force band ($A_{1}$) and the nearest sidebands  ($A_{0}$ and $A_{2}$) are given by (keeping terms up to $\ell=4$)
\begin{align}
A_0 &\approx
\frac{W}{2}(b_1+t_1)
+
\frac{3W}{16}(5W^2-4)(b_3+t_3),
\nonumber\\
A_1 &\approx
b_0
+
\frac{b_2}{4}(3W^2-2)
+
\frac{3b_4}{64}(35W^4-40W^2+8),
\label{eq:Aq_central_coefficients}\\
A_2 &\approx
\frac{W}{2}(b_1-t_1)
+
\frac{3W}{16}(5W^2-4)(b_3-t_3).
\nonumber
\end{align}
Expressions for the amplitudes $A_{q}$ for  $q=-3,\ldots,5$ are presented in the appendix \ref{app:aq_full} in terms of the $b_{\ell}$ and $t_{\ell}$,  with expressions again truncated to $\ell \leq 4$.

\begin{figure}[t]
\centering
\includegraphics[width=0.49\linewidth]{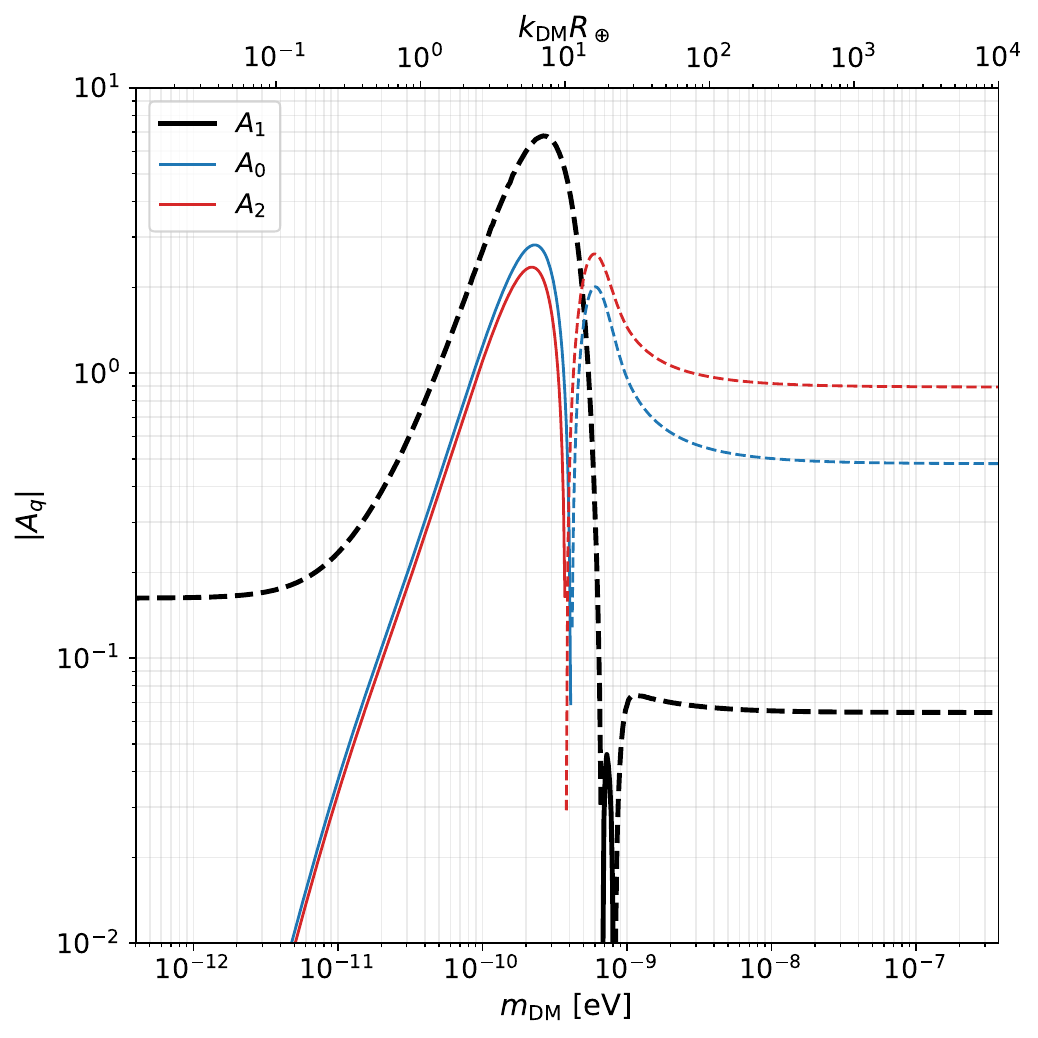}
\includegraphics[width=0.49\linewidth]{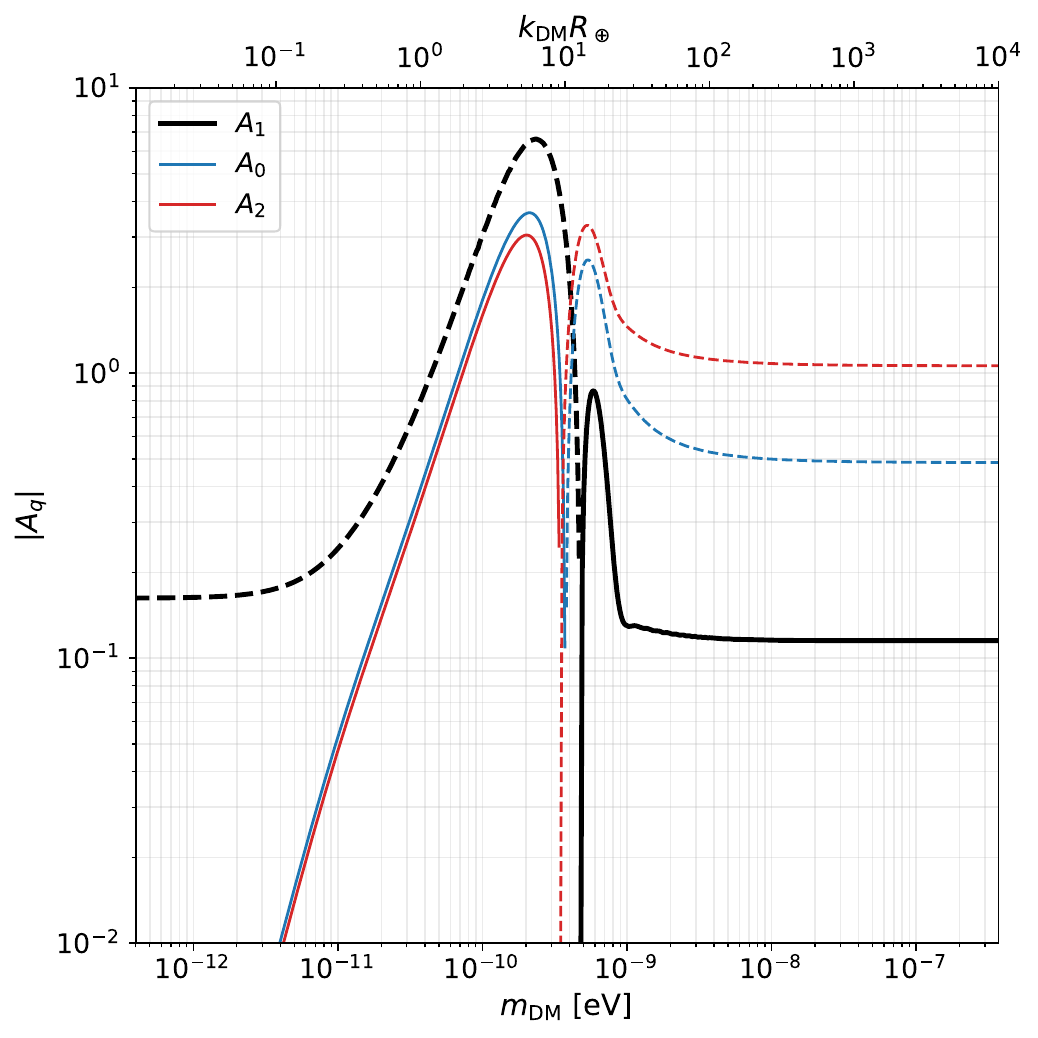}
\caption{%
Harmonic coefficients $A_q$ for $W=0.7$ (left) and $W=1.0$ (right). The lower and upper horizontal axes show $m_{\rm DM}$ and $k_{\rm DM}\Rearth$ (assuming $k_{\rm DM}\approx 10^{-3} m_{\rm DM}$), respectively, and the vertical axis shows $|A_q|$. Solid and dashed segments denote positive and negative $A_q$ values, with $q$ indicated by the legend. The relatively large magnitude of $|A_{q}|$ for $q \neq 1$ illustrates how the signal can move between the radial-force band and neighboring spin-orbit sidebands, especially at large DM mass.}
\label{fig:Aq_coefficients}
\end{figure}

Notably, these coefficients depend on the value of $W$, the magnitude of the projection of the mean DM wind direction onto the orbital plane.  This quantity evolves over time. The orbit of the MICROSCOPE had an inclination of about $98^\circ$ \cite{List2015}. Combining this with an approximate declination of the DM wind of $\delta_{\rm DM} \simeq -48^\circ$ implies $W$ ranges from $0.7\lesssim W \leq 1$ over the MICROSCOPE mission. In figure 
\ref{fig:Aq_coefficients}
we show the harmonic coefficients $A_q$ for the endpoint values $W=0.7,1$.  Comparing the two panels, we can see there are pronounced differences in the values of $|A_{q}|$ for the two $W$ values.   

Because of the dependence of the $A_{q}$ on $W$, estimating the values of $W$ sampled during the MICROSCOPE mission is important for both deriving recast constraints on $d$ and projecting the sensitivity of a dedicated analysis. In appendix \ref{app:W_estimation} we discuss in detail how the estimation is performed. Although the detailed values of the $A_q$ indeed depend on $W$, there are also some general features visible for both endpoint values of $W$. For example, the $q=0$ and $q=2$ sidebands have a larger amplitude in the high-mass regime than the $q=1$  (radial) band. Therefore, in this high-mass regime, an analysis restricted to the radial-force band alone rejects an important contribution to the DM signal, and would not give the optimal sensitivity.  

\subsection{Recast of the published MICROSCOPE constraint}
\label{sec:Recast}
We have demonstrated that taking full advantage of MICROSCOPE data will require examination of frequency bands beyond the radial-force band. However,
the published MICROSCOPE EP result does not provide a fitted amplitude for each of these sidebands. Rather, it constrains the component of the signal that overlaps with the standard EP template (the $q=1$ band). The goal of this subsection is to  derive the bound that can be obtained from this existing information alone (i.e. without additional information from the sidebands).  That is, we perform a careful recast of the published MICROSCOPE result to the DM model considered here.  Later, in section~\ref{subsec:sensitivity} we discuss the quantitative gain that might be possible using a dedicated multi-sideband analysis.
 
In our derivation of the MICROSCOPE bound in this section 
we use the circular orbit picture described above.\footnote{We expect the errors from this approximation to be generally small and subdominant to other uncertainties.}
We start with a quick summary of how MICROSCOPE limits were obtained.

The MICROSCOPE mission was optimized for an analysis in Fourier space. Despite significant gaps between science sessions, this was possible because of several design and data-processing choices~\cite{Berge:2020jgg}:
\begin{itemize}
\item Each data \emph{segment} used in the analysis spans an even integer number of orbits $N_{\rm seg,i}$. Here, $i$ labels a given segment.  After applying a discrete Fourier transform (DFT), the frequency resolution is $\Delta \Omega_{\rm seg,i} = \Omega_{\rm orb}/N_{\rm seg,i}$.  Potential signals may be placed into $\Omega_{\rm samp}/(2\Delta \Omega_{\rm seg, i})$ frequency bins, each with width $\Delta\Omega_{\rm seg, i }$. If $N_{{\rm seg},i} \simeq 10$, this can be a large number of bins, ${\mathcal O}(10^{5})$.
\item The satellite spin frequencies used for the EP segments were $\Omega_{\rm spin, V2} = (9/2) \Omega_{\rm orb}$ or $\Omega_{\rm spin, V3} = (35/2) \Omega_{\rm orb}$, resulting in an EP-violating signal appearing at frequency $\Omega_{\rm EP, V2} = (11/2) \Omega_{\rm orb}$, or $\Omega_{\rm EP, V3} = (37/2) \Omega_{\rm orb}$.
\item  After a DFT, in a given segment, the EP signal appears in the bin $N_{\rm EP, V2} = (11/2) N_{\rm seg,i}+1$ or $N_{\rm EP, V3} = (37/2) N_{\rm seg,i}+1$, because $N_{\rm seg,i}$ is an even integer. 
\end{itemize}

The MICROSCOPE data analysis uses calibrated acceleration data, masks glitches, reconstructs missing data, and then estimates the E\"otv\"os parameter $\eta_{\rm EP} \equiv 2\,\Delta a_{AB}/(a_A + a_B)$  with a frequency-domain weighted least-squares method called ADAM~\cite{Berge:2020jgg,Touboul:2022yrw}. The weight at frequency $f_k$ is set by the inverse square root of the noise power spectral density (PSD). The published global result is obtained by gathering the segment equations in Fourier space, but Ref.~\cite{Touboul:2022yrw} also shows that the global result $\eta_{\rm EP}$ and its error $\sigma_{\rm EP}$ are well-reproduced  by the inverse-variance weighted mean of the individual segment estimates,
\begin{equation}
\eta_{\rm EP}
\simeq
\frac{\sum_i \hat\eta_i/\sigma_{\eta,i}^2}{\sum_i 1/\sigma_{\eta,i}^2},
\qquad
\sigma_{\rm EP}^2
\simeq
\frac{1}{\sum_i 1/\sigma_{\eta,i}^2}.
\label{eq:published_weighted_mean}
\end{equation}
We will use the same approach to translate the MICROSCOPE result into a full-mission  bound on $d$ in terms of values derived from individual segments $d_{i}$ and their uncertainties $\sigma_{d,i}$.  This results in a central value for $d$ with uncertainty $\sigma_{d}$ given by:

\begin{equation}
d
\simeq
\frac{\sum_i  d_{i}/\sigma_{d,i}^2}{\sum_i 1/\sigma_{d,i}^2},
\qquad
\sigma_{d}^2
\simeq
\frac{1}{\sum_i 1/\sigma_{d,i}^2}.
\label{eq:published_weighted_mean_estimate}
\end{equation}

\subsubsection{Segment-level overlap}
\label{sec:inner_product}

Following standard matched-filter notation, we can compare two real waveforms $a$ and $b$ using a noise-weighted inner product~\cite{Maggiore:2007ulw}
\begin{equation}
(a|b)_i
\equiv
4\,{\rm Re}\int_0^\infty d\Omega\,
\frac{\tilde a^*(\Omega)\tilde b(\Omega)}{S_i(\Omega)} \propto
\frac{T_i}{S_i(\Omega_{\rm EP})}
\langle a(t)b(t)\rangle_i,
\label{eq:noise_weighted_inner_product}
\end{equation}
where $\langle ... \rangle_i$ denotes the time-averaging over the segment $i$, and we have anticipated our assumption that  the signal will be concentrated near the EP line. Here, $S_i$ is the noise PSD of the $i^{\rm th}$ data segment.   The same idea applies to the DFT sums used by MICROSCOPE: frequencies with larger noise PSD carry less weight. 

Neglecting systematic contributions, the MICROSCOPE analysis assumed a signal that can be represented schematically as
\begin{equation}
y(t)=\eta \, e(t)+n(t).
\label{eq:ep_segment_model}
\end{equation}
Here, $y(t) = \Gamma_{X}/g(\xi)$ is the anomalous acceleration measured along the  sensitive axis expressed in units of gravitational acceleration $g$ at a radial distance $\xi$; $n(t)$ is the noise, which is related to the noise PSD $S_{i}$ through $\langle \tilde{n}^*(\Omega) \tilde{n}(\Omega') \rangle_i = \frac{1}{2}S_i(\Omega) \delta(\Omega - \Omega')$, and $e(t)$ is the radial-like waveform:
\begin{equation}
e(t)=\cos (s(t)+a(t)),
\label{eq:ep_template}
\end{equation}
where we have assumed a circular orbit.
The published MICROSCOPE analysis reports  a best-fit amplitude $\hat\eta_i$ for the radial waveform component,  with statistical uncertainty $\sigma_{\eta,i}$. Using eq.~\eqref{eq:noise_weighted_inner_product}, we can express them as
\begin{equation}
\hat\eta_i
=
\frac{(e|y)_i}{(e|e)_i},
\qquad
\sigma_{\eta,i}^2
=
\frac{1}{(e|e)_i}.
\label{eq:eta_estimator}
\end{equation}

However, as discussed above, see eq.~(\ref{eq:aq_expansion}), the true DM signal has a  waveform that differs from the EP case:
\begin{equation}
y(t)=d\,h(t)+n(t),
\label{eq:true_signal_suboptimal}
\end{equation}
with 
\begin{equation}
h(t)
=
\frac{\NbarAB}{g(\xi)\Rearth}
\sum\limits_{q}A_{q}(t)\cos (s(t)+q\, a(t)),
\end{equation}
where we defined $\NbarAB \equiv \NAB/d$. Thus, the published EP estimator measures only the component of the true signal that overlaps with the radial-force template $e(t)$. The inferred coupling and its uncertainty are therefore determined by the projected overlap,
\begin{equation}
\hat d_{i}
=
\frac{(e|y)_i}{(e|h)_i},
\qquad
\sigma_{d,i}^{2}
=
\frac{(e|e)_i}{(e|h)_i^2}.
\label{eq:suboptimal_estimator}
\end{equation}
When $h=e$, eq.~\eqref{eq:suboptimal_estimator} reduces to eq.~\eqref{eq:eta_estimator}. Equivalently, from eq. \eqref{eq:eta_estimator} and eq. \eqref{eq:suboptimal_estimator} we have
\begin{equation}
\hat d_{i}/\hat\eta_i
=
\frac{(e|e)_i}{(e|h)_i},
\qquad
\sigma_{d,i}^{2}/\sigma_{\eta,i}^2
=
\left(\frac{(e|e)_i}{(e|h)_i}\right)^2.
\label{eq:suboptimal_estimator2}
\end{equation}
The above ratios are straightforward to evaluate. Using eq. \eqref{eq:noise_weighted_inner_product}, we get
\begin{equation}
\frac{(e|h)_i}{(e|e)_i} = \frac{\NbarAB \langle A_1 (t)\rangle_i}{g(\xi)\Rearth}.
\label{eq:mapping}
\end{equation}
Therefore, we can translate the  segment-level EP bound and error to a segment-level central value and error on $d$ as
\begin{equation}
 \hat d_{i}
=
\frac{1}{R}\frac{\hat\eta_i}{ \langle A_1 (t)\rangle_i},
\qquad
\sigma_{d,i}
=
\frac{1}{\vert R \vert}\frac{\sigma_{\eta,i}}{|\langle A_1 (t)\rangle_i|},
\label{eq:segment_recast}
\end{equation}
where we defined
\begin{equation}
R\equiv
\frac{\NbarAB}{g(\xi)\Rearth}.
\label{eq:response_factor}
\end{equation}

Combining the segments with the published statistical errors as inverse-variance weights gives
\begin{equation}
d^{q=1}
=
\frac{1}{R}\frac{\sum_i \langle A_1 (t)\rangle_i\hat\eta_i/\sigma_{\eta,i}^2}
{\sum_i \langle A_1 (t)\rangle_i^2/\sigma_{\eta,i}^2},
\qquad
\sigma_{d}^{q=1}
=
\sqrt{\frac{1}{R^2} \frac{1}{\sum_i \langle A_1 (t)\rangle_i^2/\sigma_{\eta,i}^2}}.
\label{eq:combined_recast}
\end{equation}
From our expressions for the segment-combined central value and error, we define a $2 \sigma$ bound on the coupling as $d_{*}^{q=1} \equiv \vert d^{\,q=1} \vert + 2 \sigma_{d}^{q=1}$.

The expression above shows why recasting the MICROSCOPE limits in the high-mass regime of quadratically coupled DM is subtle. Unlike the low-mass limit, where  the response in the radial-force band is approximately common to all segments, in the high-mass (directional) regime,  the relevant segment response is $R\langle A_1(t)\rangle_i$, and can vary significantly between segments. So while in the low-mass regime,
the published bound on $\eta_{\rm EP}$ can be translated by a simple overall rescaling, in the high-mass regime, the response during a given segment can be suppressed for unfavorable geometries.  This variation must be taken into account.  

\begin{figure}[t]
\centering
\includegraphics[width=0.49\linewidth]{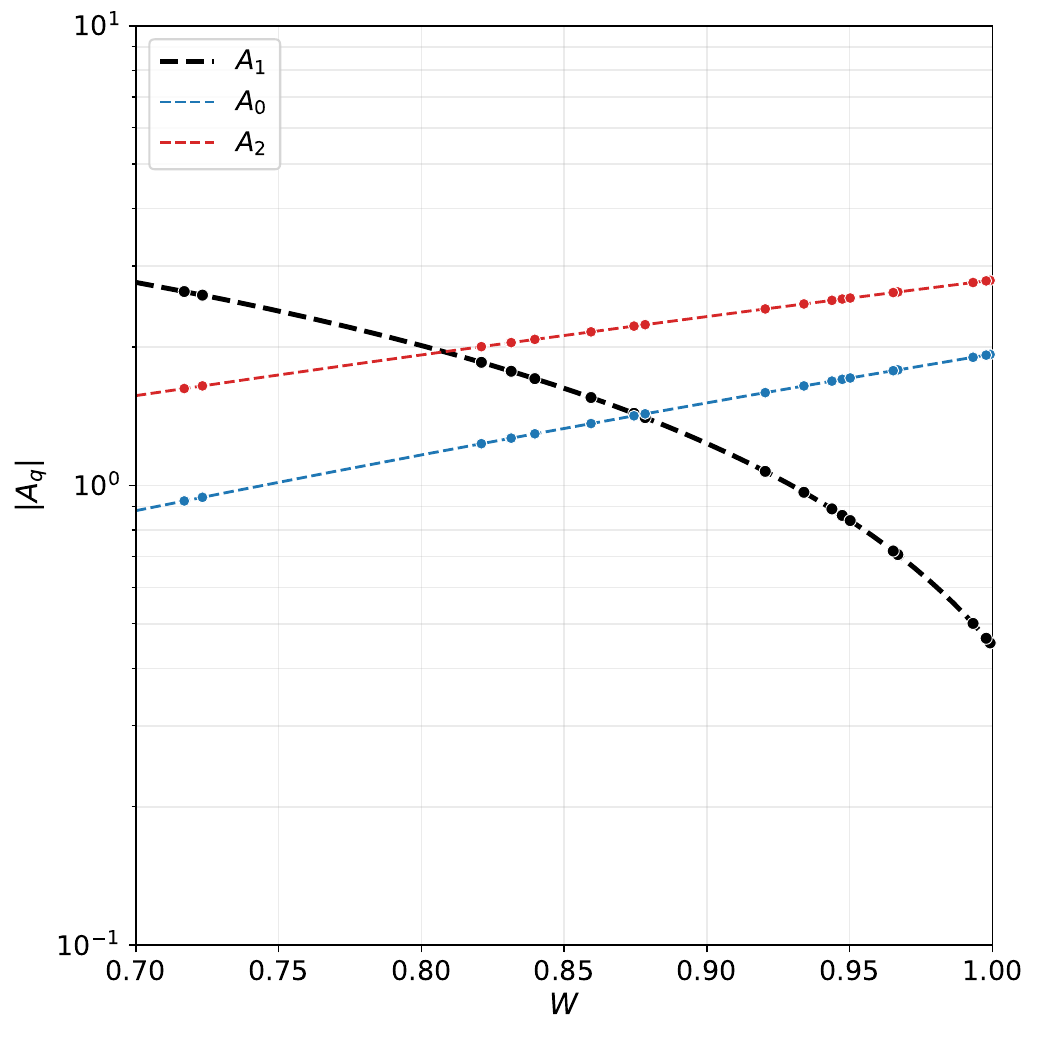}
\includegraphics[width=0.49\linewidth]{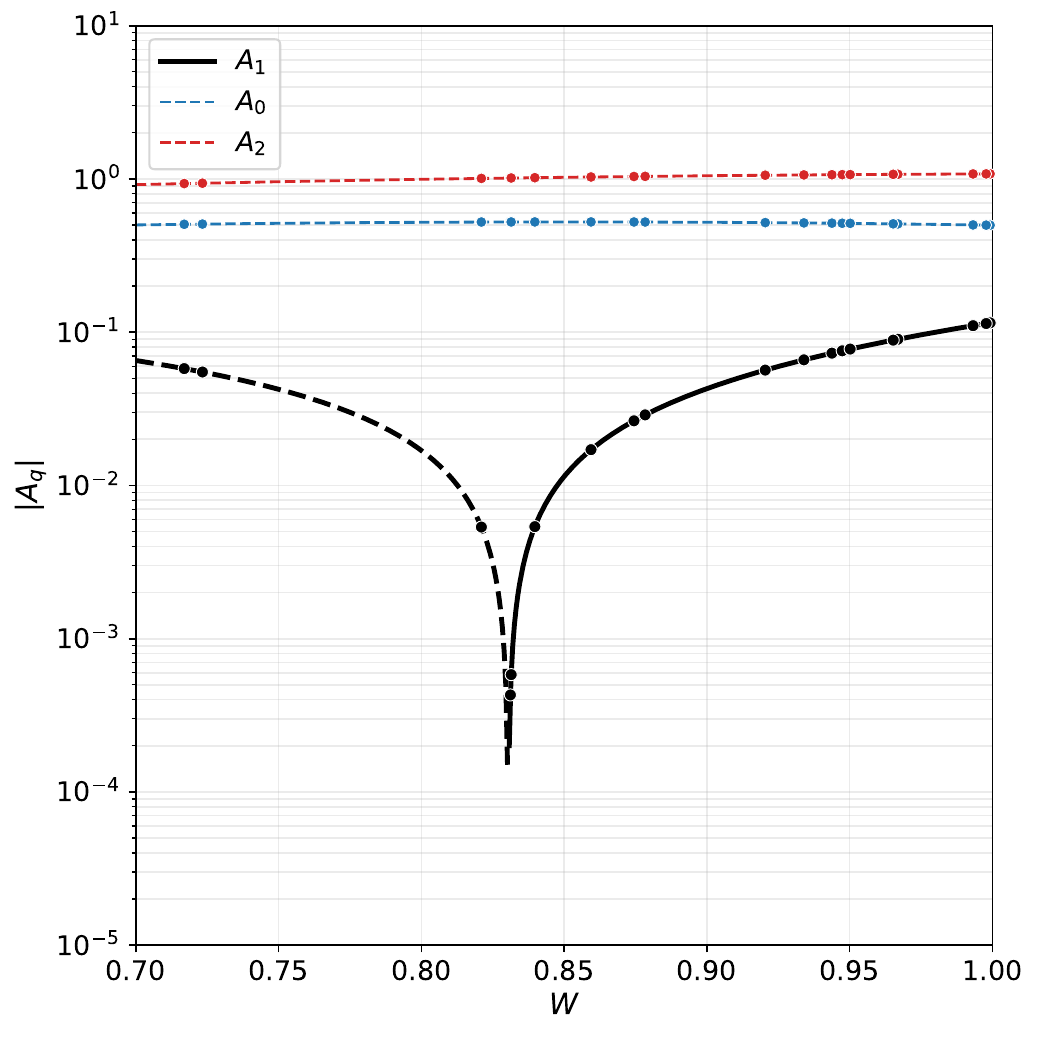}
\caption{$A_q$ values for the radial frequency band and the two nearest sidebands ($q=0,2$) as a function of $W$ for the approximate range of $W$ values sampled by MICROSCOPE. We show values for $m_{\rm DM} \approx  4.5 \times 10^{-10}$ eV  (left) and $10^{-8}$  eV (right).  
Dots represent points sampled by MICROSCOPE segments. }
\label{fig:Aq_W_sampling}
\end{figure}

An example of this effect is illustrated in figure \ref{fig:Aq_W_sampling}, where we display $|A_{q}|$, $q=0,1,2$ as a function of $W$ for two dark matter masses: $m_{\rm DM}=4.5 \times 10^{-10}$ eV (left) and  $m_{\rm DM} \approx 10^{-8}$ eV (right). Each dot represents a MICROSCOPE segment -- each segment effectively samples a particular $W$, corresponding to the relative direction of the DM wind at the time the data of that segment were taken.  For $m_{\rm DM} =4.5\times 10^{-10}$ eV (left), there is a suppression of $A_{1}$ at large $W$, which is relevant for many segments.  This results in a slight weakening of the recast bounds near this mass.  Analogously, for $m_{\rm DM} \approx 10^{-8}$ eV (right), the points near $W\simeq 0.82$ have much smaller $|A_{1}|$ and hence those segments contribute relatively little to the constraining power. We can understand the origin and location of the cancellation for this mass. As indicated in figure~\ref{fig:b_coefficients},  at this mass (and generically for $m_{\rm DM} \gtrsim 10^{-9}$ eV), $b_0$ is subdominant to $b_{2}$.  Therefore, $A_{1}$ is well approximated by the second term of eq.~(\ref{eq:Aq_central_coefficients}). The coefficient of this term vanishes at $W=\sqrt{2/3}\approx 0.82$.

\subsubsection{Screening effects}
\label{sec:Screening}

Before presenting the results of the recast, we discuss one additional physical effect.  At relatively large values of the DM-SM couplings, interactions with the satellite itself can lead to a modification of the field value near the test masses~\cite{Burrage:2025grx}.  
The region of DM masses we consider corresponds to de Broglie wavelengths much larger than the satellite size.   We therefore expect the detailed geometry of the satellite to be subleading and approximate the satellite as a constant-density sphere. 
At these long wavelengths, the field at the center of the satellite can be estimated by considering scattering off this constant density sphere.    We find that the gradient of $\langle\phi^2\rangle$, and hence the signal is suppressed relative to its unscreened background value by the factor
\begin{equation}
\label{eq:Seff}
    S(x) = \frac{2x}{\sinh(2x)} \underset{x \rightarrow \infty} \longrightarrow 4x e^{-2x}.
\end{equation}
Here, 
\begin{equation}
    x = R_{\rm sat}\sqrt{d \, \kappa^2 Q M_{\rm sat} /V_{\rm sat} } =  \left(
    \frac{d}{1.2 \times 10^{27}} \right)^{1/2}\left(
    \frac{Q}{1.7 \times 10^{-3}} \right)^{1/2} \left(\frac{M_{\rm sat}}{300 \, \rm kg} \right)^{1/2}\left(\frac{V_{\rm sat}}{2.3 \, \rm m^3} \right)^{-1/6},
    \label{eq:xScaling}
\end{equation}
where $M_{\rm sat}$ and $V_{\rm sat}=4\pi R_{\rm sat}^3/3$ denote the satellite mass and effective volume.  We take the MICROSCOPE volume corresponding to the folded configuration.\footnote{See \url{https://www.eoportal.org/satellite-missions/microscope}.}  We have chosen the fiducial value for $Q$ corresponding to the dilaton charge for aluminum for the photon coupling. 
The complete set of charges for $(e,m_{e}, g, \hat{m}, \delta m)$ is 
\begin{equation}
\label{eq:Q_Al}
Q_{\rm Al}\simeq 10^{-3} (1.7, 0.26, 1000, 81, 0.062). 
\end{equation}
From eq.~(\ref{eq:xScaling}), we can estimate at what couplings these screening effects will  be relevant. 

Using eq.~(\ref{eq:Seff}), we can account for the screening effects on the limit on $d$ by solving the equation
\begin{equation}
d_{\rm eff}(d) \equiv S(d, Q) \,  d = d_\ast, 
\end{equation}
where $d_{\ast}$ is the bound introduced in the previous subsection derived in the absence of screening.  In general, this equation yields two solutions for $d$, which we call $d_{1}$ and $d_{2}$; $d_1 < d_2$. At small couplings, $d_1$ coincides with the  unscreened limit $d_{\ast}$; as the coupling increases, it begins to deviate.  For coupling values greater than $d_{2}$, the screening from the satellite is so severe that no bound is possible.  Couplings between $d_1$ and $d_{2}$ are excluded.

While the value of $d_{1}$, i.e. the lower boundary of the excluded region, is relatively insensitive to the mass distribution within the satellite, the value of $d_{2}$ is more model dependent.  From eq.~(\ref{eq:xScaling}), $x$ depends on the volume of the satellite. This means that if the mass is more concentrated, the value of $x$ (and hence the suppression of the signal) increases. We have quantitatively investigated this effect and expect that a more realistic model of the distribution of mass and its composition within the satellite could change the value of $d_{2}$ by {$\mathcal O$}(1),  so this boundary comes with an uncertainty of this size.

\subsubsection{Results}
\begin{figure}[t!]
\centering
\includegraphics[width=0.92\linewidth]{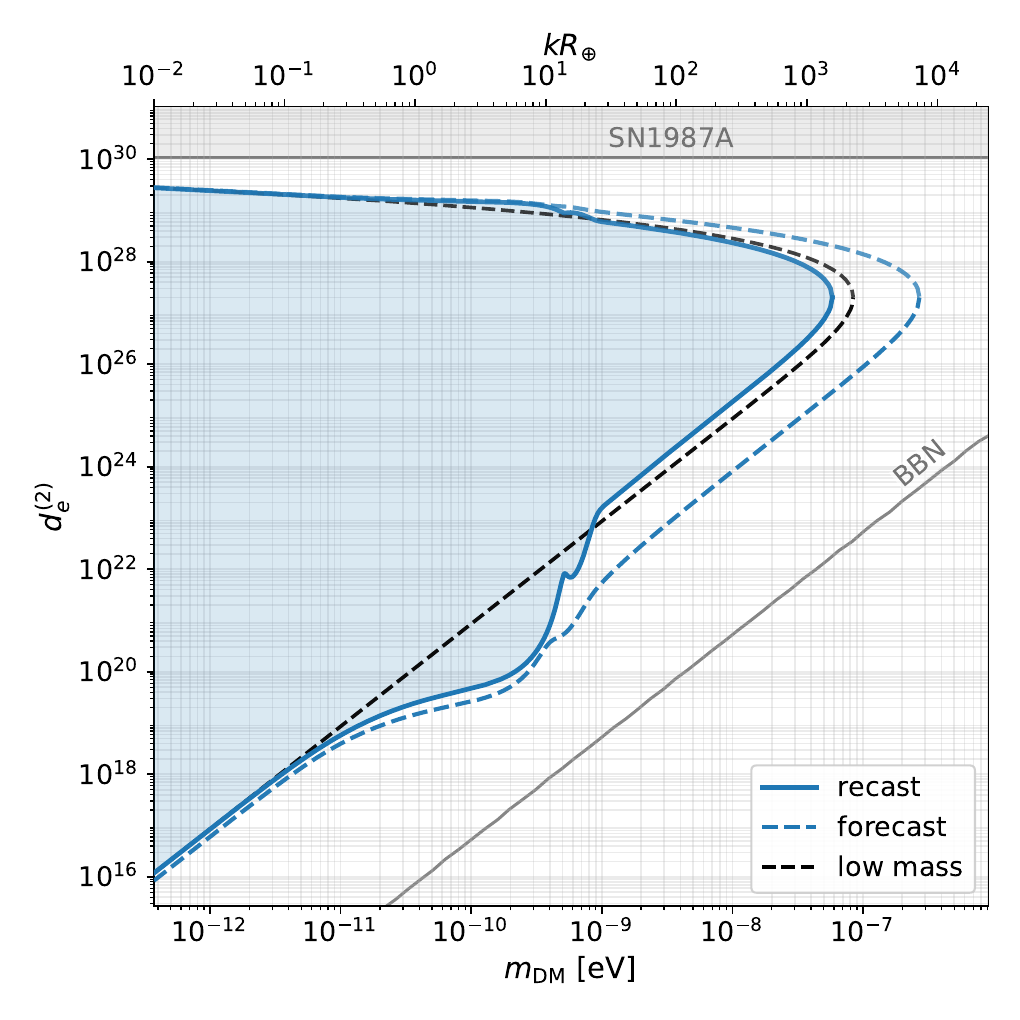}
\caption{Bound on the dark matter coupling to photons $d_e^{(2)}$ as a function of DM mass $m_{\rm DM}$. Corresponding values of $k_{\rm DM} R_{\oplus}$,  assuming $k_{\rm DM} =10^{-3} m_{\rm DM}$, are shown on the upper horizontal axis.  The solid blue line represents a $2 \sigma$ limit on $d_e^{(2)}$ obtained with the recast procedure summarized in eq. \eqref{eq:combined_recast}. To construct $\langle A_1 (t)\rangle_i$ for each data segment we use publicly available MICROSCOPE attitude data and construct the DM wind direction from the Sun’s velocity with respect to the Galactic rest frame and the Earth’s orbital velocity.
Details of this procedure are  in  appendix \ref{app:W_estimation}. The dashed black line gives the bound derived using a low-mass approximation, see text for details. The dashed blue line is the projected sensitivity for an analysis that accounts for the signal in other frequency bands, see section~\ref{subsec:sensitivity}.  The line labeled SN1987A is the bound from energy loss in supernovae from \cite{Olive:2007aj,Gan:2025nlu,Day:2023mkb}, while the line labeled BBN is a bound from Big Bang Nucleosynthesis from \cite{Bouley:2022eer}, which depends on the details of the UV completion and cosmology.}
\label{fig:de_recast}
\end{figure}

We can now present one of the primary results of this work.  In figure \ref{fig:de_recast} we show (solid blue) the $2 \sigma$ limit on the coupling of DM to photons $d^{(2)}_e$ after accounting for the time dependence of $W$  and factoring in the screening effect described above.   We also show a low-mass approximation (dashed black) that corresponds to setting $b_{0} = b_{0,LM}$, see eq.~(\ref{eq:b0_lm_raw}), with all other multipoles vanishing.  
There is a pronounced deviation  between the two (by roughly an order of magnitude) near  $m_{\rm DM} \sim 10^{-10}$ eV. This is a direct reflection of $b_{0}$ 
deviating from its low-mass value near these masses, as can be seen in the left panel of figure~\ref{fig:b_coefficients}.  The geometric effects highlighted in the discussion of figure~\ref{fig:Aq_W_sampling} are reflected in the weakening of the bound for $m_{\rm DM} \approx  4 \times 10^{-10}$ eV. As mentioned above, we expect the upper boundary, above which screening effects prevent a bound, to be uncertain at the {$\mathcal O(1)$} level due to incomplete modeling of the mass distribution within the satellite. 
Also, in the derivation described here, we have ignored the effects of the atmosphere, simply modeling the Earth as a hard sphere. 
In appendix~\ref{app:Atmosphere}, we quantify the associated atmospheric systematic using atmosphere-modified partial waves. This comparison indicates that the lower boundary of the excluded region is relatively robust, while the upper, screening-limited boundary can receive $\mathcal O(1)$ corrections.
The recast bounds for other couplings $d^{(2)}_{X}$, $X\in \{g, m_{e}, \hat{m},\delta m\}$ are presented in appendix \ref{app:OtherCouplings}. 

\subsection{Potential gain from a dedicated multi-sideband analysis}
\label{subsec:sensitivity}

The goal of this subsection is to demonstrate the power of a dedicated multi-harmonic analysis.  We show that it could lead to significantly stronger limits than the above recast that was based only on the existing radial-force constraint.  
The key difference from the recast is that we now assume a matched filter for the full DM signal, rather than projecting the signal onto the published EP template. Under these assumptions, 
\begin{equation}
\sigma_{d,i}^2
=
\frac{1}{(h|h)_i} = \sigma_{\eta,i}^2 \frac{(e|e)_i}{(h|h)_i}.
\label{eq:sigma_d_estimator_harmonic}
\end{equation}
Here, the $q\neq1$ sidebands are also localized in DFT bins, and are well displaced from the radial-force band (by $(q-1)N_{\rm seg,i}$ bins). We assume that the noise across all sidebands is approximately the same and equal to the noise in the EP bin, 
$S_i(\Omega_{q,i})\simeq S_i(\Omega_{{\rm EP},i})$. Should the true noise in the nearest sidebands substantially differ, this would impact the precise gain from this analysis.
Under this assumption, the above expression simplifies to
\begin{equation}
\sigma_{d,i}^2
= \frac{1}{R^2} \frac{\sigma_{\eta,i}^2}{ \sum_q \langle A_q^2(t)\rangle_i}.
\label{eq:sigma_d_estimator_harmonic2}
\end{equation}
Combining segments with inverse-variance weights gives
\begin{equation}
\sigma_{d}^{\text{all }q}
=
\sqrt{\frac{1}{R^2} \frac{1}{\sum_{i,q} \langle A_q^2 (t)\rangle_i/\sigma_{\eta,i}^2}}.
\label{eq:multi_sigma}
\end{equation}

In figure~\ref{fig:de_recast}, we show the potential improvement in the inferred bound on the DM-photon coupling $d_{e}^{(2)}$ obtained by using the full multi-sideband signal. In contrast to the recast, for this sensitivity projection we set the central value to zero, $d_{*}^{\rm all \,q} \equiv  2\, \sigma_{d}^{\rm all \,q}$.  This explains the residual difference between the recast and projected curves at low masses. As expected, the sensitivity gain is especially significant at higher DM masses where the anisotropy in the dark matter profile induced by the Earth is most pronounced. The central result of this figure is that for $m_{\rm DM} \gtrsim 4 \times 10^{-10} \, \rm eV$, a dedicated analysis could improve the coupling reach by over an order of magnitude relative to a radial-force recast. Similar improvements are possible in other DM couplings as shown in appendix \ref{app:OtherCouplings}.  It is also notable that the dip in sensitivity present at $m_{\rm DM} \sim 5 \times 10^{-10}$ eV is largely smoothed out in the sensitivity projection.  This dip corresponds to a case where $A_{1}$ is small (in part due to a large number of segments with unfavorable geometries, analogous to figure \ref{fig:Aq_W_sampling}).  However, when $A_{1}$ is small, typically there is a relatively large contribution to $A_{2}$ which can be used to constrain the coupling.

\section{Conclusion}
\label{sec:outlook}

In this work we have laid the foundation for a future analysis of MICROSCOPE data in the mass regime where the scalar-induced force is anisotropic. 

Building on the field-profile formalism of  Ref.~\cite{Brzeminski:2026rox}, we constructed the corresponding angular decomposition of the force. The result is naturally organized in terms of radial $b_\ell$ and polar $t_\ell$ force coefficients, with the polar axis set by the mean DM momentum $\khat$. For a fixed DM mass, these coefficients determine the relative multipole content of the signal, leaving the overall coupling normalization as the parameter to be constrained.

We then translated this force into the MICROSCOPE observable by projecting it onto the time-dependent measurement axis.  
A purely radial force produces the usual monochromatic radial-force modulation at a frequency $\Omega_{\rm spin}+\Omega_{\rm orb}$ due to  the rotation and orbital motion of the satellite. 
By contrast, the anisotropic coefficients $b_{\ell>0}$ and $t_{\ell>0}$ populate additional resolved sidebands of the form $\Omega_{\rm spin}+q\, \Omega_{\rm orb}$, with integer $q$. 

We used this template in two ways. First, we derived a recast of the published MICROSCOPE EP result by retaining only the component of the DM signal that overlaps with the standard $q=1$ radial-force template. In the directional regime this recast is not a simple global rescaling of the published EP bound. The relevant response depends on the segment-averaged geometry through the projection of the DM wind onto the orbital and measurement axes. Second, we estimated the sensitivity of an idealized dedicated analysis that uses the full multi-sideband signal.

Neighboring sidebands can carry a substantial fraction of the modeled signal power and, after averaging over the actual MICROSCOPE segment geometry, the EP ($q=1$) band can be strongly suppressed relative to the full signal. In the high-mass regime, taking advantage of these sidebands leads to a projected improvement of more than an order of magnitude. This result strongly motivates a dedicated MICROSCOPE data analysis. 

Such an analysis should also refine the environmental modeling that controls the screening-limited edge of the reach. For atmospheric effects, we have taken a first step in appendix~\ref{app:Atmosphere} by solving the atmosphere-modified partial-wave problem for a simplified exponential density profile and an NRLMSIS-derived spherically averaged profile~\cite{NRLMSIS}. We find that the hard-surface approximation is robust at the lower boundary of the recast exclusion region, while atmospheric corrections can become large near the upper, screening-limited boundary. A full reanalysis should therefore combine the multi-sideband template with an atmosphere-modified field profile as well as a more realistic treatment of the MICROSCOPE satellite geometry, especially when determining the largest couplings and highest masses that can be excluded.
\\

\noindent
{\bf Note Added:} While this work was being completed, we became aware of overlapping work in preparation \cite{TienTien}.

\section*{Acknowledgments}
D.B. and A.P. are supported by
the Department of Energy under grant number DE-SC0007859. D.B. is supported in part by a Leinweber postdoctoral fellowship.

\appendix

\section{Dilatonic Charges}
\label{app:dilaton_charges}

For an element with atomic number $Z$ and mass
number $A$, we use \cite{Damour:2010rm,Damour:2010rp}
\bea
Q_e(A,Z)
&=&
10^{-4}\left[
-1.4+8.2\frac{Z}{A}
+7.7\frac{Z(Z-1)}{A^{4/3}}
\right],\\
Q_{m_e}(A,Z)
&=&
5.5\times10^{-4}\frac{Z}{A},\\
Q_{\hat{m}}(A,Z)
&=&
0.093-\frac{0.036}{A^{1/3}}
-0.020\left(\frac{A-2Z}{A}\right)^2
-1.4\times10^{-4}\frac{Z(Z-1)}{A^{4/3}},\\
Q_{\delta m}(A,Z)
&=&
0.0017\frac{A-2Z}{A},\\
Q_g(A,Z)
&=&
1.093-Q_e(A,Z)-Q_{m_e}(A,Z)-Q_{\hat{m}}(A,Z).
\label{eq:dilatonic_charges}
\eea

\section{Harmonic coefficients}
\label{app:aq_full}

Here we present the $A_{q}$ coefficients of eq.~\eqref{eq:aq_expansion} for $q=-3, \ldots, 5$, truncating each expression at  $\ell\leq4$.  We give the results for fixed $W$.
They are:
\begin{align}
A_{-3} &= \frac{35W^4}{128}(b_4+4t_4),
\nonumber \\
A_{-2} &= \frac{5W^3}{16}(b_3+3t_3),
\nonumber\\
A_{-1} &=
\frac{3W^2}{8}(b_2+2t_2)
+
\frac{5W^2}{32}(7W^2-6)(b_4+2t_4),
\nonumber\\
A_0 &=
\frac{W}{2}(b_1+t_1)
+
\frac{3W}{16}(5W^2-4)(b_3+t_3),
\nonumber\\
A_1 &=
b_0
+
\frac{b_2}{4}(3W^2-2)
+
\frac{3b_4}{64}(35W^4-40W^2+8),
\label{eq:Aq_coefficients}\\
A_2 &=
\frac{W}{2}(b_1-t_1)
+
\frac{3W}{16}(5W^2-4)(b_3-t_3),
\nonumber\\
A_3 &=
\frac{3W^2}{8}(b_2-2t_2)
+
\frac{5W^2}{32}(7W^2-6)(b_4-2t_4),
\nonumber\\
A_4 &= \frac{5W^3}{16}(b_3-3t_3),
\nonumber \\
A_5 &= \frac{35W^4}{128}(b_4-4t_4).
\nonumber
\end{align}

\section{Master integrals for force coefficients}
\label{app:master_integrals}

The force coefficients follow from the same one-dimensional integrals that determine the field multipoles, see Ref.~\cite{Brzeminski:2026rox}. Throughout this work we assume that the DM field is given by
\bea
 \phi(\Vec{x},t) = \int d^3k \sqrt{f_{\rm BMB}(\Vec{k})} \Tilde{\phi}(\Vec{x},t;\Vec{k}), 
 \eea
where $\Tilde{\phi}(\Vec{x},t;\Vec{k})$ is a solution of eq.~\eqref{eq:KG} given an incoming plane wave of momentum $\Vec{k}$, and $f_{\rm BMB}(\Vec{k})$ is the boosted Maxwell-Boltzmann distribution
\bea
\label{eq:BoostedMBapp}
f_{\rm BMB}(\Vec{k}) = 
\l \frac{1}{2 \pi \sigma_k^2} \r^{3/2} \exp\l -\frac{(\Vec{k}-\Vec{k_{\rm DM}})^2}{2 \sigma_k^2} \r.
\eea
The plane wave solution can be decomposed into time- and space-dependent parts
\bea
\Tilde{\phi}(\Vec{x},t;\Vec{k}) \equiv \text{Re}(e^{i \omega t +i\chi_{\Vec{k}}} \psi(\Vec{x};\Vec{k})).
\eea
For the gradient force we are interested in understanding the angular structure of the field gradient $\langle \nabla \phi^2\rangle$.
This is the gradient of the profile derived in Ref.~\cite{Brzeminski:2026rox}. We will begin with a review of the computation of the profile.

\subsection{Field profile review}

After averaging over the rapid oscillations and random phases, the field profile depends only on the spatial part of the field solution $\psi$

\bea
\label{eq:averagedphiApp}
 \langle \phi^2(\Vec{x},t) \rangle &=& \int d^3p \, d^3k \sqrt{f_{\rm BMB}(\bm p)f_{\rm BMB}({\bm k})} \langle  \Tilde{\phi}(\Vec{x},t;\Vec{p}) \Tilde{\phi}(\Vec{x},t;\Vec{k})\rangle \\
&=& \int d^3k f_{\rm BMB}({\bm k}) \vert \psi(\Vec{x},\Vec{k}) \vert^2 .
\eea
In the second line, we used the fact that terms with different momenta $\bm k \neq \bm p$ average to zero because they have independent phases.
Since the boundary condition is spherically symmetric, we decompose the solution into partial waves
\begin{equation}
\psi=
\abs{\psi_0}
\sum_{\ell}(2\ell+1)i^\ell
P_\ell(\cos\theta)F_\ell(k,\xi),
\label{eq:psi_partial}
\end{equation}
where $\theta$ is the angle between the position $r$ and the momentum $k$. In  spherical coordinates where the polar axis is specified by $\khat$, we have
\begin{equation}
\cos\theta
=
\hat{\bm{k}}\cdot\rhat
=
\sin\theta_k\cos\varphi_k\sin\theta_r
+
\cos\theta_k\cos\theta_r.
\label{eq:appdotproduct}
\end{equation}
By azimuthal symmetry ($f_{\rm BMB}$ is independent of $\varphi_k$), we can set $\varphi_r = 0$, which was used to simplify the above form.

For the field profile, we are interested in an expression for $\vert \psi \vert^2$ in terms of multipoles.    
Using angular momentum addition properties to collect terms of the same total angular momentum, we get
\begin{equation}
\abs{\psi}^2=
\abs{\psi_0}^2
\sum_m c_m(k,\xi)P_m(\cos\theta),
\label{eq:psi_sq_legendre}
\end{equation}
with
\begin{equation}
c_m(k,\xi)=
\sum_{\ell,\ell'}
(2\ell+1)(2\ell'+1)i^{\ell-\ell'}
C^r_{\ell,\ell',m}
F_\ell(k,\xi)F_{\ell'}^*(k,\xi).
\label{eq:c_m_def}
\end{equation}
Here, $C^r_{\ell\ell'L}$ can be represented in terms of the Wigner-3j symbol
\begin{equation}
  C^r_{\ell\ell'L}
  =
  \frac{2L+1}{2}
  \int_{-1}^{1}\dd z\,
  P_\ell(z)P_{\ell'}(z)P_L(z)
  =
  (2L+1)
  \begin{pmatrix}
    \ell & \ell' & L\\
    0 & 0 & 0
  \end{pmatrix}^{\!2}.
  \label{eq:Cr_def}
\end{equation}
The expectation value $\langle \phi^2({\bf x},t)\rangle$, which requires an integration against the BMB distribution, can be neatly organized into radial and angular parts.  We write:
\bea
\label{eq:averagedphiDecomp}
 \langle \phi^2(\Vec{x},t) \rangle &=& \abs{\psi_0}^2\int dk \, k^2   
\sum_m c_m(k,\xi)\, \int d\cos \theta_k\, f_{\rm BMB}(k, \theta_k) \, \int d\varphi_k P_m(\cos\theta).
\eea
Since only $\cos \theta$ depends on $\varphi_k$, see eq.~(\ref{eq:dotproduct}), we perform this integration first, which gives
\bea
\int\limits_{0}^{2\pi} d \varphi_k P_l (\cos \theta) = 2 \pi P_l (\cos \theta_k) P_l (\cos \theta_r).
\eea
The result has an important consequence: the remaining angular integral (over $\cos \theta_{k}$), reduces to a calculation of an angular momentum projection of the BMB distribution, which can be performed analytically. We define it as
\begin{equation}
f_m(k)\equiv 
\int_{-1}^{1}
\dd(\cos\theta_k)\,
f_{\rm BMB}(k,\cos\theta_k)P_m(\cos\theta_k).
\label{eq:f_m_def}
\end{equation} 
The entire profile can be compactly expressed as 
\begin{equation}
    \langle \phi^2(\Vec{x},t) \rangle \equiv \frac{\rho_{\rm DM}}{m_{\rm DM}^2} \sum_m a_m P_m (\cos \theta_r),
\end{equation}
where the  multipole coefficients $a_m$ that describe the field are given by 
\begin{equation}
a_m(\xi)=
2\pi\int_0^\infty\dd{k}\,k^2
c_m(k,\xi)f_m(k),
\label{eq:appendix_master}
\end{equation}
with the $c_{m}$ defined in eq.~(\ref{eq:c_m_def}).
These integrals can be numerically evaluated for the multipoles of interest.

\subsection{Gradient force}
 A directional description of the gradient force follows directly from the above expression. From the gradient in spherical coordinates
\bea
\nabla = \frac{1}{R_\oplus}\l \partial_\xi \rhat + \frac{1}{\xi} \partial_{\theta }\thetahat + \frac{1}{\xi \sin \theta} \partial_\varphi \phihat \r,
\eea
we get 
\bea
    -\nabla \langle \phi^2 \rangle &=& -\frac{\abs{\psi_0}^2}{R_{\oplus}}\sum_m \lbrace  (\partial_\xi  a_m) P_m (\cos \theta_r) \rhat + \frac{1}{\xi}  a_m (\partial_{\theta} P_m (\cos \theta_r) )\thetahat \rbrace,\\
    &=& \frac{\abs{\psi_0}^2}{R_{\oplus}} \sum_m \left\lbrace   b_m P_m (\cos \theta_r) \, \rhat +  t_m  P_m^1 (\cos \theta_r) \, \thetahat \right\rbrace.
\eea
In the above, we have  defined
\begin{equation}
b_m(\xi) \equiv
-2\pi\int_0^\infty\dd{k}\,k^2
\frac{\partial c_m(k,\xi)}{\partial \xi}
f_m(k),
\qquad
t_m(\xi) \equiv
-\frac{a_m(\xi)}{\xi}.
\label{eq:raw_master}
\end{equation}
The derivative that acts on $c_m$ will act on the partial wave terms $F_{\ell}(k,\xi)$ and $F_{\ell'}(k,\xi)$, see eq.~(\ref{eq:c_m_def}), with result: 
\begin{equation}
\frac{\partial c_m(k,\xi)}{\partial \xi}=
\sum_{\ell,\ell'}
(2\ell+1)(2\ell'+1)i^{\ell-\ell'}
C^r_{\ell,\ell',m} \l
\frac{\partial F_\ell(k,\xi)}{\partial \xi}F_{\ell'}^*(k,\xi) + \frac{\partial F_{\ell'}^*(k,\xi)}{\partial \xi}F_{\ell}(k,\xi)\r.
\label{eq:c_m_deriv}
\end{equation}

\subsection{Lowest Multipoles}

We now present explicit integral expressions for the lowest-order coefficients.  For the monopole,
\begin{equation}
a_0(\xi)=
\frac{1}{\sqrt{2\pi}k_{\rm DM}\sigma_k}
\int_0^\infty\dd{k}\,k
\left[
e^{-\frac{(k-k_{\rm DM})^2}{2\sigma_k^2}}
-
e^{-\frac{(k+k_{\rm DM})^2}{2\sigma_k^2}}
\right]
\sum_{\ell}(2\ell+1)\abs{F_\ell(k,\xi)}^2,
\label{eq:a0_explicit}
\end{equation}

For the dipole part of the field profile we have:
\begin{align}
a_1(\xi)
=&
-\frac{6}{\sqrt{2\pi}k_{\rm DM}\sigma_k}
\int_0^\infty\dd{k}\,
\left[
\left(k+\frac{\sigma_k^2}{k_{\rm DM}}\right)
e^{-\frac{(k+k_{\rm DM})^2}{2\sigma_k^2}}
+
\left(k-\frac{\sigma_k^2}{k_{\rm DM}}\right)
e^{-\frac{(k-k_{\rm DM})^2}{2\sigma_k^2}}
\right]
\nonumber\\
&\hspace{2.0cm}\times
\sum_{\ell}(\ell+1)
\text{Im}\!\left[
F_{\ell+1}(k,\xi)F_\ell^*(k,\xi)
\right].
\label{eq:a1_explicit}
\end{align}
These expressions immediately yield formulas for the monopole ($b_0$) and dipole ($b_1$, $t_1$) gradient-force coefficients via
eq.~(\ref{eq:raw_master}).

\section{$W(t)$ reconstruction}\label{app:W_estimation}

The goal of this appendix is to explain how the geometric factor $W$ is estimated for each segment.  As introduced in section~\ref{sec:microscope_template}, this parameter determines the amplitude of 
$\mu(t) = \rhat (t) \cdot \khat (t) = W(t)\cos \left[\phi(t)-\chi_w \right]$ modulation during the orbit.

We find it convenient to express $W(t)$ in terms of the DM wind direction $\khat(t)$ and the orbital angular momentum direction $\nhat (t)$: 
\bea
\label{eq:appWoft}
  W^2(t)=1-\left(\khat(t)\cdot \nhat(t)\right)^2 .
\eea
This form naturally divides the computation into two parts.  First, we need to know where the dark-matter wind points in an inertial frame at each sampled time. Second, because the spin axis of the satellite is approximately parallel to the orbital axis\footnote{We verified this assumption by transforming the gravitational-acceleration data and confirming that it is suppressed along $+\hat X_{\rm sat}$, see App.~\ref{sec:appChecks}.}, we can
extract the orbital angular momentum direction by finding the spin axis of the satellite in the same frame. We use the J2000
frame for this comparison. By J2000 we mean the standard inertial celestial reference
frame tied to the epoch J2000.0; in practice it is the common frame in which the
ephemeris vectors and the satellite attitude directions are expressed.
In the publicly available MICROSCOPE data the spin axis can be found via the satellite attitude:
the satellite spin axis is parallel to the $\hat X_{\rm sat}$ axis. Here, we follow the conventions of  \cite{Robert:2020ddm}.  The satellite frame and the instrument frame are not identical.  The relationship is $X_{\rm inst} \approx -Z_{\rm sat}$,  $Y_{\rm inst} \approx X_{\rm sat}$, $Z_{\rm inst} \approx -Y_{\rm sat}$.  The sensitive axis denoted as ${\bf \hat{X}}$ in the main text corresponds to $X_{\rm inst}$.    
Since $W^2$ depends only on $(\khat\cdot\nhat)^2$, the sign of the normal is irrelevant for the reconstruction of $W$. Operationally, we choose
\bea
  \nhat(t)\simeq \xsat^{\rm J2000}(t),
\eea
where $\xsat^{\rm J2000}(t)$ is obtained from the attitude data, see appendix~\ref{app:attitude} for details.

\subsection{Dark matter wind direction}

We define the wind direction using the velocity of the Earth through the Galactic
dark-matter frame.  The adopted solar Galactocentric velocity is first expressed in the
J2000 frame.  
We then add the Earth's orbital velocity relative to
the Sun,
\bea
  \mathbf v_{\rm \oplus, GC}(t)
  =
  \mathbf v_{\odot,{\rm GC}}^{\rm J2000}
  +
  \mathbf v_{\oplus,{\rm orb}}(t).
\eea
For the solar term we use
\[
  \mathbf v_{\odot,{\rm GC}}^{\rm J2000}
  =
  (114,\ -122,\ 181)\ {\rm km\,s^{-1}}.
\]
This value is obtained from the Astropy Galactocentric-frame $v4.0$ solar-motion
convention, transformed from Galactic Cartesian components into the J2000/ICRS
Cartesian basis.\footnote{Astropy Galactocentric-frame defaults:
\url{https://docs.astropy.org/en/stable/api/astropy.coordinates.galactocentric_frame_defaults.html}.
For the ICRS/J2000 convention, see the IERS ICRS documentation:
\url{https://www.iers.org/IERS/EN/DataProducts/ICRS/icrs}.}
The time-dependent term $\mathbf v_{\oplus,{\rm orb}}(t)$ is evaluated from
solar-system ephemerides as the Earth's velocity relative to the Sun. We use Astropy's solar-system position and velocity tools for this step; the relevant documentation is the Astropy barycentric position/velocity interface.\footnote{Astropy
solar-system position and velocity documentation:
\url{https://docs.astropy.org/en/stable/api/astropy.coordinates.get_body_barycentric.html}.
For the reference-frame conventions used by JPL Horizons, see
\url{https://ssd.jpl.nasa.gov/horizons/manual.html}.}
The direction used in the $W$ calculation is
\bea
  \khat(t)
  =
  -
  \frac{\mathbf v_{\rm \oplus,GC}(t)}
  {|\mathbf v_{\rm \oplus,GC}(t)|}.
\eea
The ephemeris is evaluated at a representative set of times inside each segment.

\subsection{Attitude and spin axis extraction}
\label{app:attitude}

The satellite attitude is provided in the MICROSCOPE data as quaternions. A quaternion $q(t) = (\cos \frac{\alpha}{2}, \hat{T} \sin \frac{\alpha}{2} )$ is simply a compact way of representing the rotation from the satellite frame to the inertial J2000 frame.
It provides information about the rotation axis $\hat{T}$ and the angle of rotation $\alpha$. 
Applying this rotation to a vector in the
satellite frame tells us where that vector points in J2000 at the corresponding time.

Starting from  $\hat{X}_{\rm sat}$ in the satellite frame,  
  $\xsat^{\rm sat}=(1,0,0)$,
we rotate it into the J2000 frame using the attitude quaternion,
\bea
\label{eq:appnoft}
  \nhat(t)\simeq \xsat^{\rm J2000}(t)
  =
  R_{\rm sat\to J2000}(q(t))\,\xsat^{\rm sat} =
\begin{pmatrix}
\cos\alpha+(1-\cos\alpha)(T_1)^2\\
(1-\cos\alpha)T_1 T_2+\sin\alpha\, T_3\\
(1-\cos\alpha)T_1 T_3-\sin\alpha\, T_2
\end{pmatrix} .
\eea
Here, $T_i$ represents a component of the rotation axis vector $\hat{T}$. 

\subsection{Instantaneous $W$ and segment moments}

Once $\khat(t)$ and $\nhat(t)$ are known, we may evaluate $W(t)$ via eq.~(\ref{eq:appWoft}).
We find 
$\langle W\rangle$, $\langle W^2\rangle$, and higher moments such as
$\langle W^3\rangle$ and $\langle W^4\rangle$.  These moments  enter the corresponding segment-averaged sideband amplitudes $\langle A_q \rangle_i$, $\langle A_q^2 \rangle_i$.  In this sense, table \ref{tab:w-segment-summary} below is the compact record of the
geometric part of the calculation: it shows the sampled range of $W$ and the two
lowest moments used by the recast and the forecast.

\begin{table}[htbp]
\centering
\small
\caption{Segment-level $W$ summaries from the matched wind and attitude sampling.}
\label{tab:w-segment-summary}
\begin{tabular}{lrrrr}
\toprule
Segment & $\min W$ & $\max W$ & $\langle W\rangle$ & $\langle W^2\rangle$ \\
\midrule
210 & 0.8275 & 0.8350 & 0.8312 & 0.6909 \\
212 & 0.8347 & 0.8452 & 0.8398 & 0.7052 \\
218 & 0.8613 & 0.8880 & 0.8745 & 0.7648 \\
234 & 0.9097 & 0.9309 & 0.9205 & 0.8473 \\
236 & 0.9310 & 0.9561 & 0.9438 & 0.8908 \\
238 & 0.9561 & 0.9768 & 0.9669 & 0.9349 \\
252 & 0.9883 & 0.9972 & 0.9932 & 0.9865 \\
254 & 0.9972 & 1.0000 & 0.9992 & 0.9984 \\
256 & 0.9944 & 0.9999 & 0.9978 & 0.9957 \\
326-1 & 0.8466 & 0.8721 & 0.8594 & 0.7387 \\
326-2 & 0.8719 & 0.8848 & 0.8784 & 0.7716 \\
358 & 0.9348 & 0.9593 & 0.9474 & 0.8976 \\
402 & 0.9634 & 0.9672 & 0.9653 & 0.9318 \\
404 & 0.9367 & 0.9631 & 0.9502 & 0.9029 \\
406 & 0.9314 & 0.9365 & 0.9340 & 0.8724 \\
438 & 0.8288 & 0.8342 & 0.8315 & 0.6913 \\
442 & 0.8189 & 0.8233 & 0.8210 & 0.6741 \\
748 & 0.7126 & 0.7214 & 0.7170 & 0.5140 \\
750 & 0.7216 & 0.7251 & 0.7233 & 0.5232 \\
\bottomrule
\end{tabular}
\end{table}

\subsection{Orbit-axis and convention checks}
\label{sec:appChecks}
In the above steps we relied on $X_{\rm sat}$ as a proxy for the orbit-normal direction and estimated the DM wind direction and orbital axis in the J2000 frame.

We double-check that $X_{\rm sat}$ is indeed a good proxy for this orbit-normal direction  by evaluating the component of the gravitational acceleration along $X_{\rm sat}$.  The gravitational acceleration direction should be a proxy for the radial direction.  We confirm that the component of the gravitational acceleration along $X_{\rm sat}$ is indeed strongly suppressed compared to its total amplitude.

To verify that we agree on conventions regarding the J2000 frame, we construct an independent expectation for the orbital normal from the
declared description of the orbit as a dawn-dusk orbit with $98^\circ$ inclination\footnote{For the numerical orbital inclination we use the public CelesTrak General Perturbations element set for MICROSCOPE, NORAD catalog ID 41457 / COSPAR 2016-025B, retrieved 22 June 2026. The record at epoch 2026-06-19T23:45:47 UTC gives INCLINATION = 98.2438 deg, which we round to $i \simeq 98^\circ$.}  \cite{MICROSCOPE:2022doy}.  This gives a simple
geometric model for the angular-momentum direction $\nhat_{\rm model}(t)$ evaluated in the J2000 frame, against which the attitude-derived
$+\hat X_{\rm sat}^{J2000}$ direction of eq.~(\ref{eq:appnoft}) can be compared. We find that for any data segment the 
$\nhat_{\rm model}(t) \cdot \hat X_{\rm sat}^{\rm J2000}(t) \approx 1 $ up to $10^{-3}$ level. This provides a consistency check of our frame and attitude conventions.

\section{Results for other couplings}
\label{app:OtherCouplings}
\begin{figure} 
\includegraphics[width=0.49\linewidth]{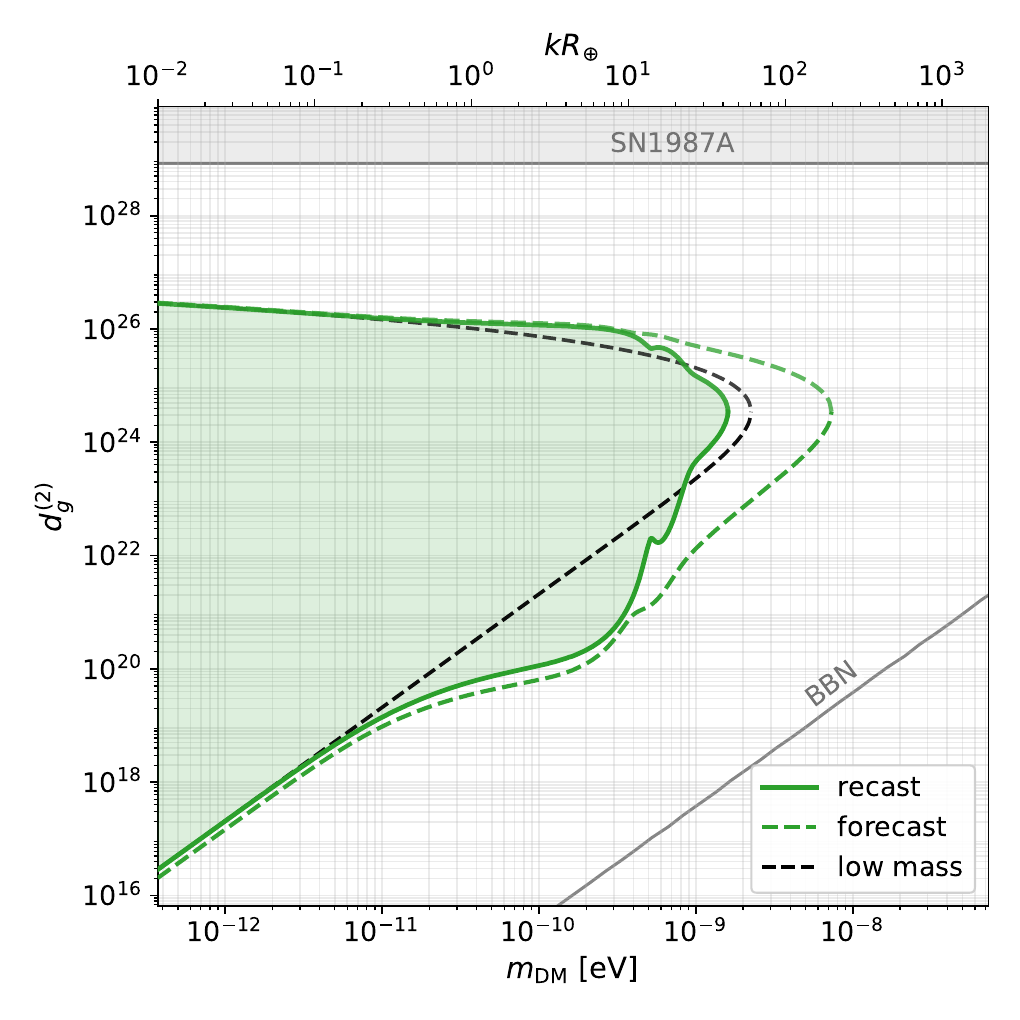}
\includegraphics[width=0.49\linewidth]{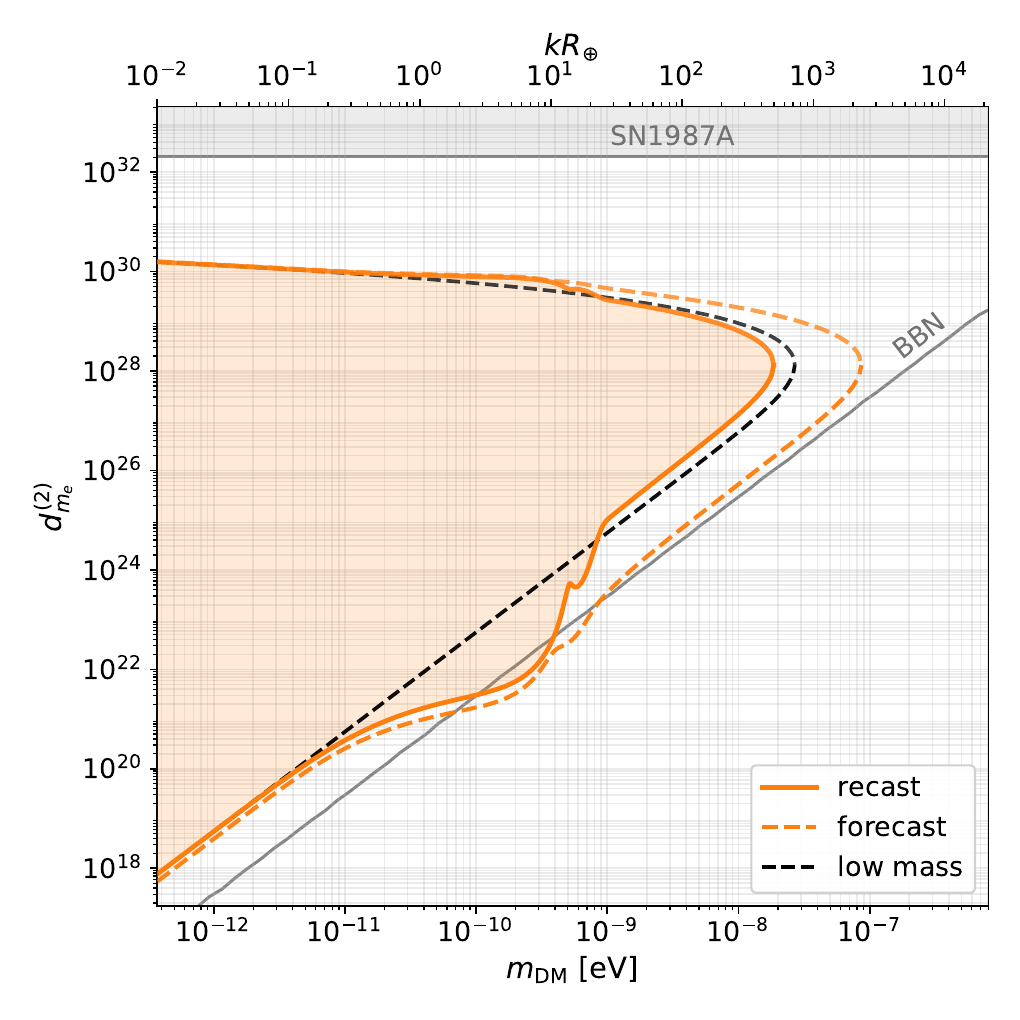}
\includegraphics[width=0.49\linewidth]{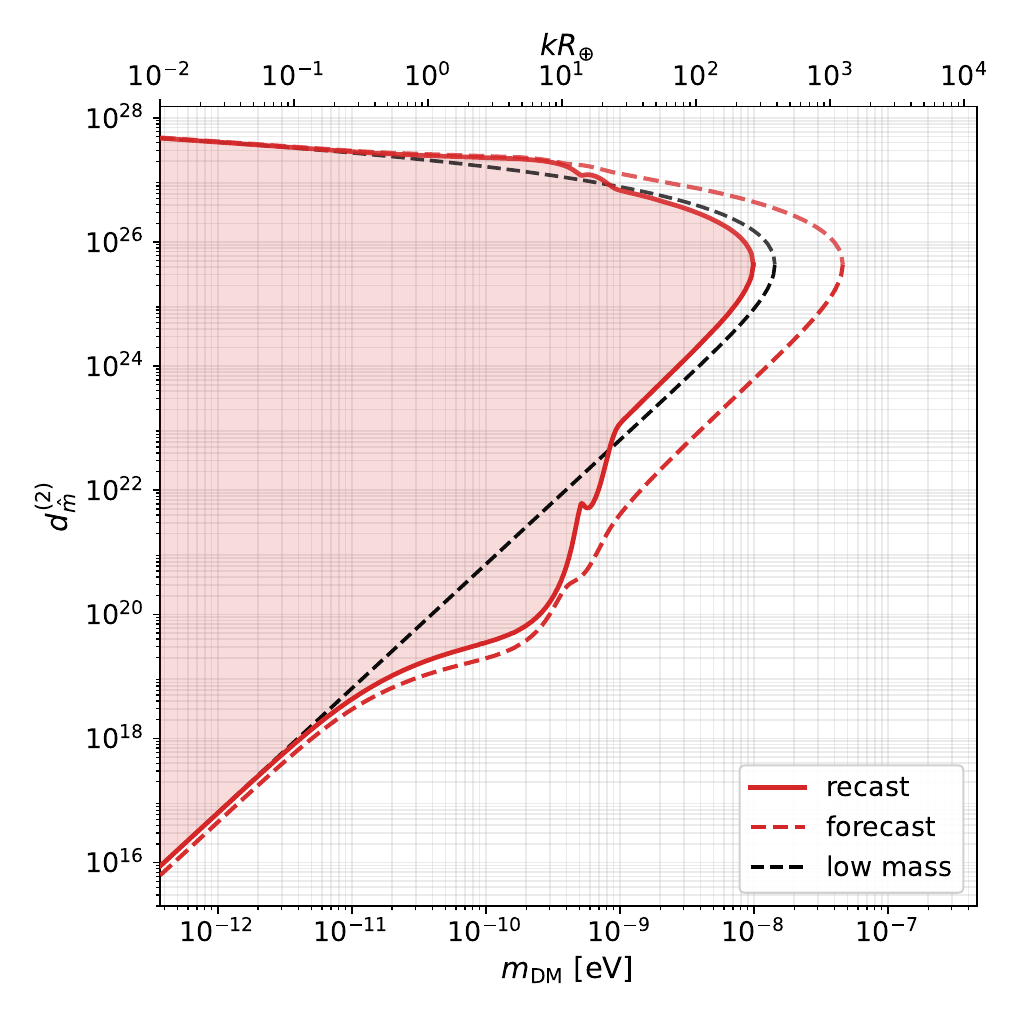}
\includegraphics[width=0.49\linewidth]{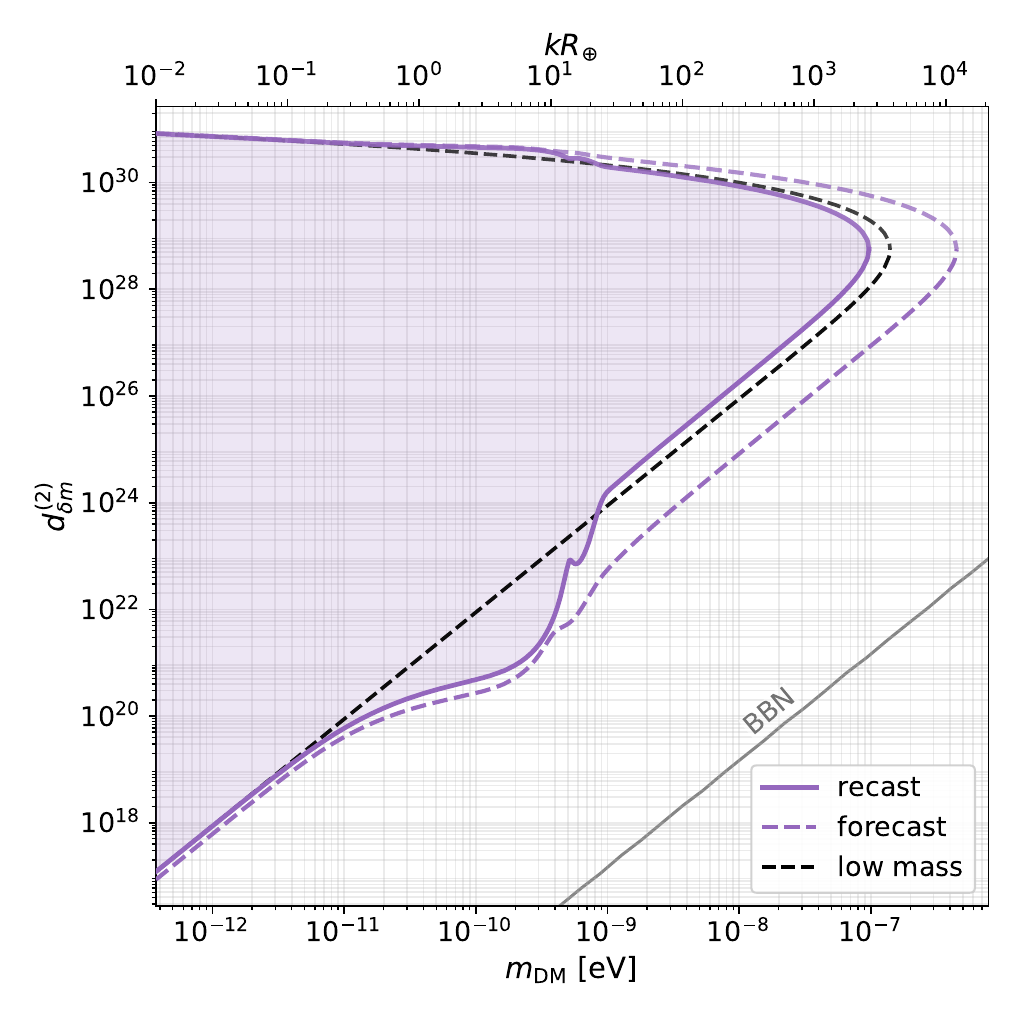}

\caption{ Recast bounds on  $d^{(2)}_{X}$, $X\in \{g, m_{e}, \hat{m},\delta m\}$ as a function of DM mass $m_{\rm DM}$. Corresponding values of
$k_{\rm DM} R_{\oplus}$ assuming $k_{\rm DM} = 10^{-3} m_{\rm DM}$  are shown on the upper horizontal axis,  see caption of figure~\ref{fig:de_recast} for more details. }
\label{fig:otherRecasts}
\end{figure}

In figure~\ref{fig:otherRecasts} we present recasts of the published MICROSCOPE EP result for the other couplings $d_X^{(2)}$, $X \in \{g, m_{e}, \hat{m}, \delta m \}$, following the methodology of section~\ref{sec:Recast}. 
These are presented as solid colored curves. The low-mass approximations are shown as dashed-black curves. Also shown are colored dashed curves corresponding to projected sensitivities if a dedicated analysis taking advantage of the anisotropy of the dark matter profile were to be performed, see section~\ref{subsec:sensitivity}.  As discussed in the main text, see also appendix \ref{app:Atmosphere}, we expect there to be ${\mathcal O} (1)$ uncertainties in the upper part of the exclusion regions of the recast. These reflect simplified modeling of the  distribution of mass within the satellite and the atmospheric density profile.  In the regime where screening is unimportant (i.e. the lower boundary of the excluded region), the ratio between these bounds and the one for $d_{e}^{(2)}$ shown in the text is given by the ratio of the associated differences of dilaton charges $\Delta Q^{AB}$, see eq.~(\ref{eq:deltaQ}).  Note also that the bound on $d_{g}^{(2)}$ does not extend to as high masses as the other couplings.  This is because the associated dilaton charge of the satellite is largest for this coupling. 
Screening effects can therefore become relevant at lower values of the coupling. 

\section{Atmosphere-modified partial waves}
\label{app:Atmosphere}

To obtain results for the recast and forecasts we used a field profile obtained by imposing a
Dirichlet boundary condition at the surface of the Earth. This gives a
``hard-surface baseline" wherein the Earth is treated as an opaque object but ignores the atmosphere.

All matter effects
for $r>\Rearth$ are set to zero. This hard-sphere approximation is useful because it isolates the dominant scattering
effect of the Earth. However, it should not be reliable for
arbitrarily large coupling -- for sufficiently large coupling, the relatively dilute atmosphere generates a non-negligible potential.

In this appendix we solve the scalar partial-wave problem in the presence of a spherically averaged atmospheric potential. We then use the resulting atmosphere-modified partial waves to estimate the corresponding corrections to the force coefficients entering the MICROSCOPE recast. Concretely, we compare the $b_\ell$ and $t_\ell$ coefficients obtained in the hard-surface baseline to those obtained after propagating each partial wave through the atmospheric potential.
 
In the presence of the atmospheric potential  
$V_{\rm atm}(h) = d_X^{(2)}Q_X(h)\kappa^2\rho_{\rm atm}(h)$, the radial wavefunction $u_\ell$ for the $\ell^{\rm th}$ partial wave  satisfies
\begin{equation}
  u_{\ell}''(x)
  +
  \frac{2}{x}u_{\ell}'(x)
  +
  \left[
    1
    -
    \frac{\ell(\ell+1)}{x^2}
    -
    \frac{V_{\rm atm}(h)}{k^2}
  \right]u_{\ell}(x)=0 ,
  \label{eq:atmosphere_radial}
\end{equation}
where we defined
\begin{equation}
  x=kr,\qquad K=k\Rearth,\qquad h=r-\Rearth .
\end{equation}
We impose the same Dirichlet boundary condition at the Earth's surface used in the hard-surface calculation: 
\begin{equation}
u_\ell(K)=0,
\end{equation}
but now solve outward in the presence of $V_{\rm atm}$ until a region is reached where the atmospheric contribution is negligible: $V_{\rm atm}(h)/k^2\ll1$. The numerical solution is then matched onto
the free exterior basis,
\begin{equation}
u_\ell(x)=A_\ell j_\ell(x)+B_\ell n_\ell(x),
\end{equation}
which determines the atmosphere-corrected $F_\ell$, 
\bea\label{eq:F_atm}
F_\ell(x) = j_\ell(x) -\frac{i B_\ell}{A_\ell + i B_\ell} h_\ell(x).
\eea

Since the atmospheric potential is, to a good approximation, spherically symmetric, it
does not mix different $\ell$-modes. We can therefore keep the same
decomposition into $b_\ell$ and $t_\ell$ coefficients.
Operationally, it means substituting the $F_\ell$ of eq.~\eqref{eq:psi_partial} with eq.~\eqref{eq:F_atm} for each $\ell$. While this increases the numerical cost, it is still feasible to evaluate $b_\ell$ and $t_\ell$ up to $k R_\oplus \sim 100$, which covers most of the range relevant for the recast. Although this coverage is not sufficient for a full atmosphere-corrected recast, it provides a useful estimate of the possible size of the atmospheric systematic by comparing these $b_{\ell}/t_{\ell}$ to those evaluated in the hard-surface approximation.

\subsection{Atmospheric modeling}
\label{sec:atmModel}
Before presenting the results of this comparison, we must first specify the potential $V_{\rm atm}(h)$ in eq.~(\ref{eq:atmosphere_radial}).  
This potential is determined by the atmospheric density profile.  
We consider two models for the atmosphere to help gauge the sensitivity to the precise profile that is assumed.   The first is a simple exponential model 
\begin{equation}
  \rho_{\rm exp}(h)=\rho_{0}\exp(-h/H),
  \qquad h=r-\Rearth ,
  \label{eq:vatm_exp}
\end{equation}
with $\rho_{\rm 0}=1.2\times10^{-3}\,{\rm g}/{\rm cm}^3$ and $H=6.9\,{\rm km}$. This profile is suitable only as an approximation to the lower atmosphere, roughly up to $100$ km. It should therefore be most reliable for coupling values for which the region satisfying $\frac{V_{\rm atm}(h)}{k^2} \gtrsim1$ is contained below $h \sim 100$ km. For stronger coupling values we expect this model to be insufficient. 

We also use a profile derived from NRLMSIS~\cite{NRLMSIS}, an empirical atmosphere model that provides altitude-, location-, time-, solar-activity-, and geomagnetic-activity-dependent atmospheric densities. 
We construct a spherically symmetric radial density profile by averaging the NRLMSIS density over geographic coordinates and over the times corresponding to the MICROSCOPE segment intervals, using the corresponding solar and geomagnetic inputs.
Schematically,
\begin{equation}
  \rho_{\rm NRLMSIS}(h)=\langle\rho_{\rm NRLMSIS}(h,\Omega,t)\rangle_{\Omega,t}.
  \label{eq:rho_NRLMSIS}
\end{equation}  
Comparing these two profiles allows us to gauge the sensitivity of the atmosphere-modified partial waves, and hence of the force coefficients, to the assumed atmospheric density profile.
\subsection{Atmospheric corrections}
\begin{figure}[t]
  \centering
  \includegraphics[width=0.49\textwidth]{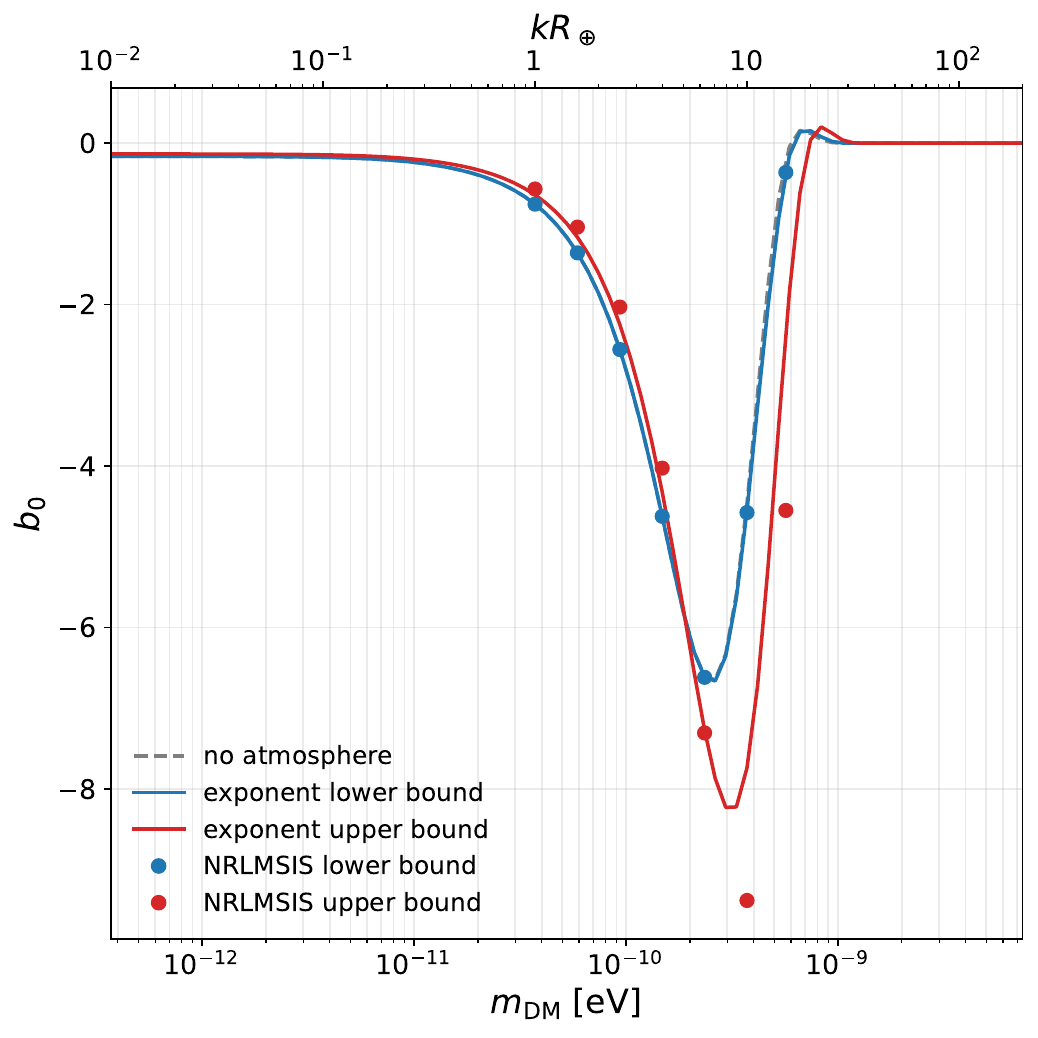}
\includegraphics[width=0.49\textwidth]{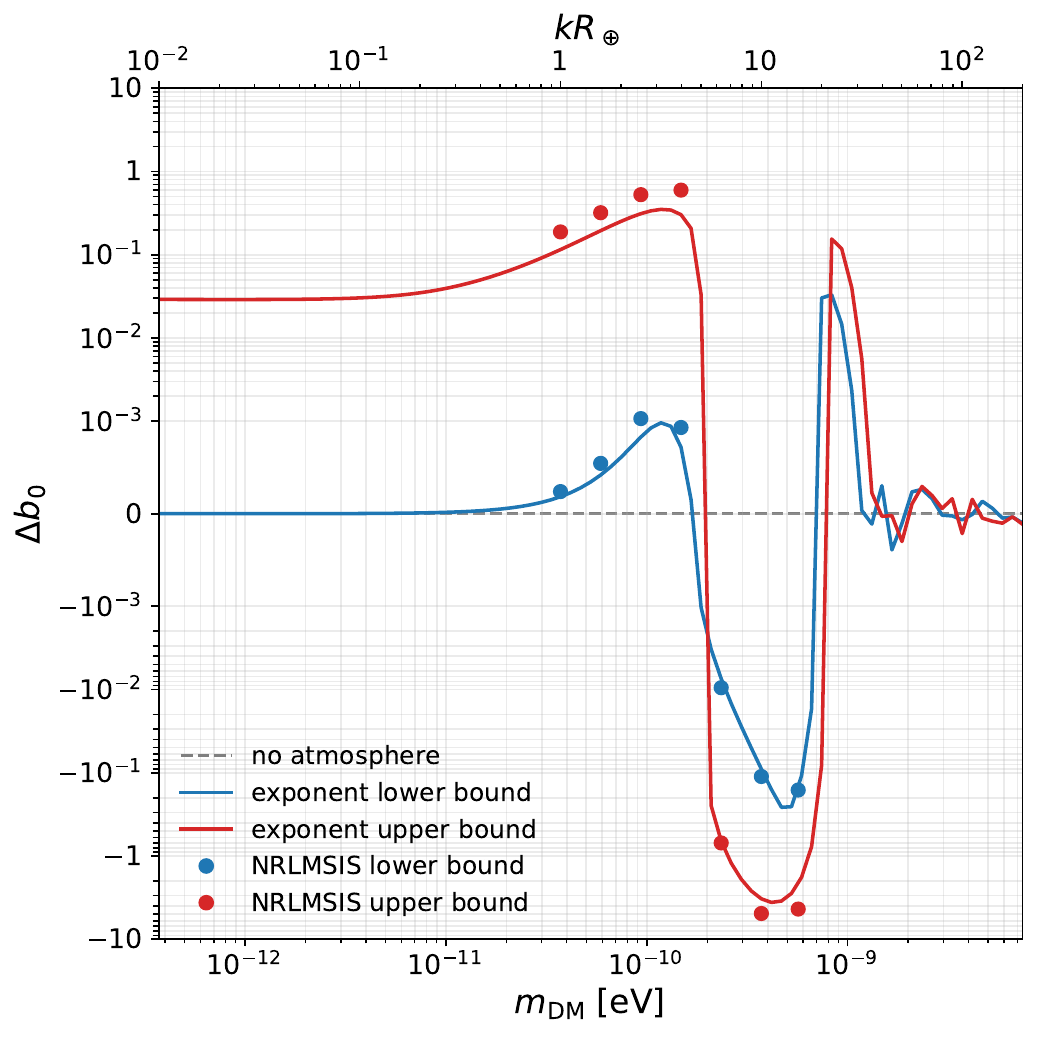}
\includegraphics[width=0.48\textwidth]{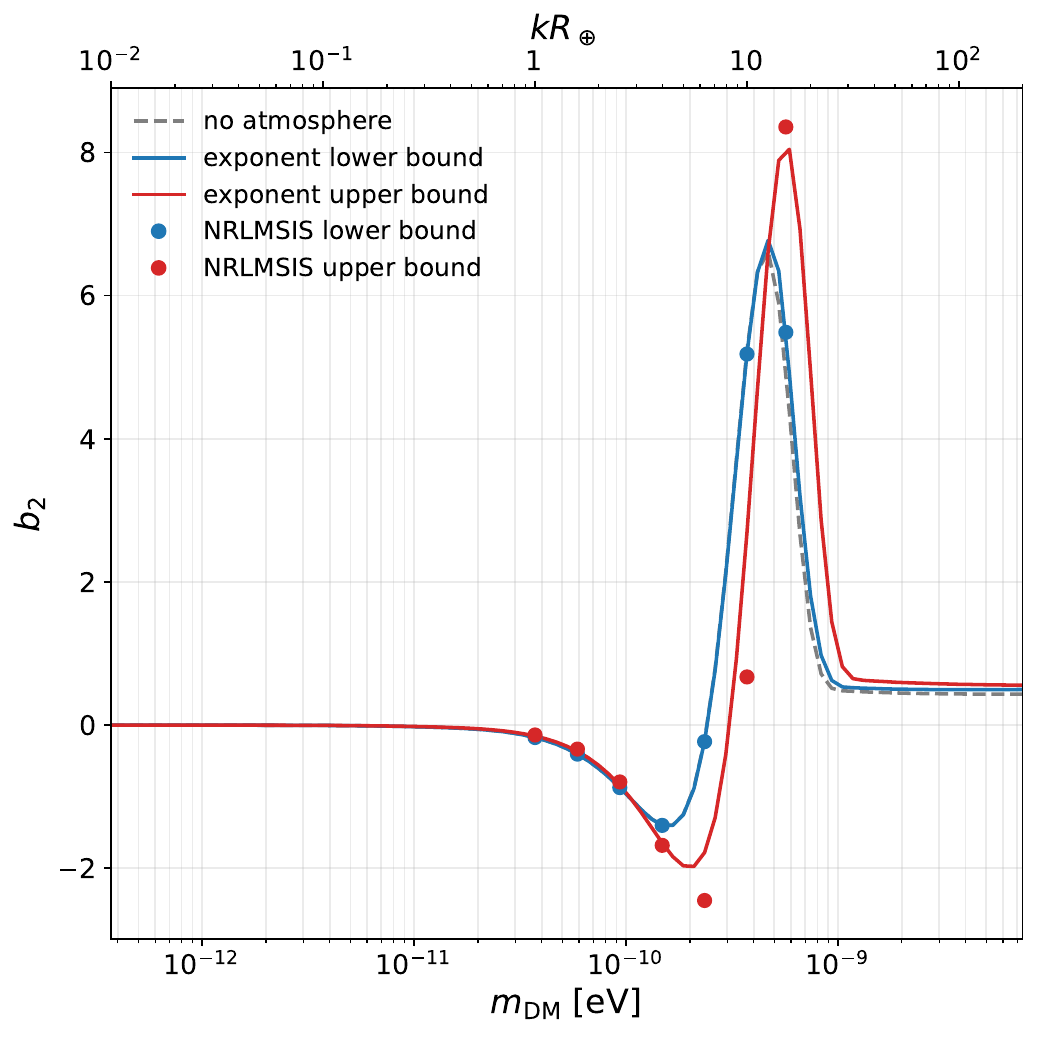}
\includegraphics[width=0.48\textwidth]{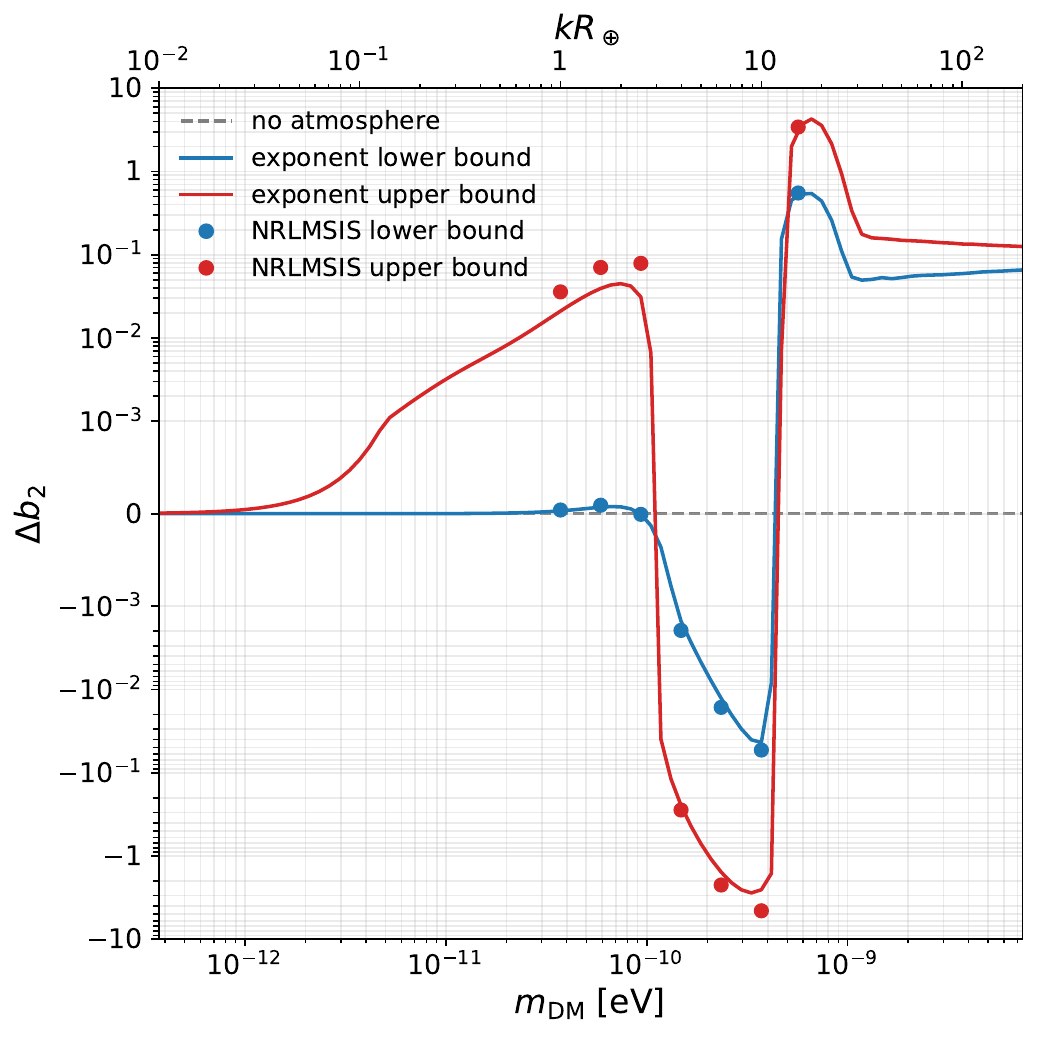}
  \caption{Comparison of $b_0$ (upper panels) and $b_2$ (lower panels) coefficient values obtained assuming no atmosphere (dashed line), single-exponential atmospheric model (colored lines), and the spherically averaged NRLMSIS-derived profile (dots). The blue (red) colored lines and dots correspond to $d_e^{(2)}$ values on the lower (upper) recast boundary in figure \ref{fig:de_recast}.
  Left panels show the values of $b_0$ and $b_2$ coefficients, while the right panels show residuals  relative to the no-atmosphere baseline. }
  \label{fig:b0}
\end{figure}

The panels in figure~\ref{fig:b0} show results for $b_0$ and $b_2$, which are most relevant for the recast bounds. The dashed gray curve
shows the hard-surface result used in the baseline calculation. Once the effects of the atmosphere are taken into account, and we are no longer strictly in the limit of hard-sphere scattering, the DM profile depends on the value of the DM-SM coupling. The blue (red) curve shows the values for $b_0$ and $b_{2}$ including the exponential atmospheric potential, with $d_e^{(2)}$ chosen at each mass to match the lower (upper) recast boundary. The dots represent results obtained using the NRLMSIS-derived profile, with the same coloring convention. 
In the left panels, we show values of the coefficients; the right panels show residuals with respect to the calculation neglecting the atmospheric contribution. 

We see that for values of $d_{e}^{(2)}$ corresponding to the lower boundary of the recast,  the $b_{0,2}$ (shown in blue) are close to the hard-surface baseline for both atmospheric models. 
For the $b_0$ coefficient, the
relative change stays at sub $10^{-3}$ level, except for $kR_\oplus \sim 10$, where the coefficient changes quickly with $k R_\oplus$. Even then the relative change does not exceed $30\%$. The $b_2$ coefficient exhibits similar behavior in the $kR_{\oplus} \sim 10$ regime.  However, when $k R_{\oplus} \sim 100$, the correction to $b_2$ remains at the few-percent level.  

The upper recast line result (red) is qualitatively different. Particularly around $k R_\oplus\sim10$, both atmosphere models produce order-one changes in both
$b_0$ and $b_2$.  In that region the minima and peaks of these coefficients are visibly shifted and more pronounced. Thus, any relative shift of the curve is exacerbated by the fact that the function is changing rapidly. For $b_2$, as on the lower recast boundary, the correction does not vanish at larger $k R_\oplus$. Around $k R_\oplus\sim 100$ it remains at the $O(10)\%$ level. The upper recast boundary probes stronger couplings.  Because the simple exponential model tends to underestimate densities at large altitudes, it may miss effects at these large couplings.  
This, in part, explains the difference  between the simple exponential and the NRLMSIS-derived coefficients.  
This also suggests that the correction due to the atmosphere at $k R_{\oplus} \sim 100$ may be larger than the single-exponential estimate for the upper-bound.

From these explorations, we  conclude that the hard-surface calculation appears robust at the lower recast boundary, potentially up to localized features in the partial-wave coefficients. However, at the upper recast boundary, atmospheric effects can become large enough that a future atmosphere-corrected analysis should either use a more realistic atmospheric model or include this effect as a systematic uncertainty.

\bibliographystyle{JHEP}
\bibliography{biblio}

\end{document}